\documentclass[]{aa}
\usepackage[utf8]{inputenc}
\usepackage[T1]{fontenc}
\usepackage{txfonts}
\usepackage{graphicx}
\usepackage[colorlinks=true, linkcolor=blue, filecolor=blue, citecolor=blue, urlcolor=blue]{hyperref}    
\usepackage{natbib}
\bibpunct{(}{)}{;}{a}{}{,} 
\makeatletter
\renewcommand*\aa@pageof{, page \thepage{} of \pageref*{LastPage}}
\makeatother

\begin{document}
\title{Differential reddening in 48 globular clusters:\\ 
An end to the quest for the intracluster medium}
\titlerunning{Differential reddening in globular clusters}
\author{
        E.~Pancino\inst{\ref{oaa}},
        A.~Zocchi\inst{\ref{vienna}},
        M.~Rainer\inst{\ref{oaa},\ref{oami}},
        M.~Monaci\inst{\ref{mmm}},
        D.~Massari\inst{\ref{oas}},
        M.~Monelli\inst{\ref{iac}},
        L.~K.~Hunt\inst{\ref{oaa}},
        L.~Monaco\inst{\ref{andresbello}},
        C.~E.~Mart\'\i nez-V\'azquez\inst{\ref{gem}},
        N.~Sanna\inst{\ref{oaa}},
        S.~Bianchi\inst{\ref{oaa}},
        P.~B.~Stetson\inst{\ref{dao}}
        }
\authorrunning{E.~Pancino et al.}

\institute{INAF - Osservatorio Astrofisico di Arcetri, Largo E. Fermi 5, I-50125 Firenze, Italy\label{oaa}
\and Department of Astrophysics, University of Vienna, T\"urkenschanzstrasse 17, A-1180 Vienna, Austria\label{vienna}
\and INAF - Osservatorio Astronomico di Brera, Via E. Bianchi 46, I-23807 Merate (LC), Italy\label{oami}
\and Dipartimento di Fisica, Universit\`a di Pisa, Largo Bruno Pontecorvo 3, I-56127 Pisa, Italy \label{mmm}
\and INAF - Osservatorio di Astrofisica e Scienza dello Spazio di Bologna, Via Gobetti 93/3, I-40129 Bologna, Italy\label{oas}
\and  Instituto de Astrof\'\i sica de Canarias, Calle V\'\i a L\'actea, E-38205 La Laguna, Tenerife, Spain\label{iac}
\and Universidad Andres Bello, Facultad de Ciencias Exactas, Departamento de Ciencias F\'isicas - Instituto de Astrofisica, Autopista Concepci\'on-Talcahuano, 7100, Talcahuano, Chile\label{andresbello}
\and Gemini Observatory/NSF's NOIRLab, 670 N. A'ohoku Pl., Hilo, HI 96720, USA\label{gem}
\and Herzberg Astronomy and Astrophysics, National Research Council, 5071 West Saanich Road, Victoria, British Columbia V9E 2E7, Canada\label{dao}
}
   
\date{Received: \today}

\abstract{For decades, it has been theorized that a tenuous but detectable intracluster medium should be present in globular clusters, which is continuously replenished by the gas and dust ejected by bright giants and periodically cleared by interactions with the Galactic disk. However, dedicated searches, especially in infrared and radio wavelengths, have returned mostly upper limits, which are lower than theoretical expectations by several orders of magnitude. We profited from recent wide-field photometry for 48 Galactic globular clusters to compute high-resolution maps of differential reddening, which can be used to correct any photometric catalog in these areas for reddening variations. Using 3D reddening maps from the literature, we evaluated the amount of foreground extinction. This allowed us to estimate the masses of the intracluster medium in our sample clusters, with an accuracy of one order of magnitude. Our estimates agree with the few available literature detections and with theoretical expectations. Because the discrepancy between observations and expectations only concerns literature upper limits, we explored possible reasons why they could be underestimated and we show that two recent discoveries can explain the discrepancy. The first is the recent discovery that the intracluster medium in 47\,Tuc is not centrally concentrated. This is also supported by our maps, which in the majority of cases do not show a central reddening concentration. The second is the discovery that the dust in metal-poor ([Fe/H]\,$\lesssim$\,$-$1\,dex) globular clusters is dominated by iron grains rather than silicates, which undermines previous dust mass estimates from observed upper limits. We conclude that current evidence, including our maps, does not contradict theoretical expectations and the problem of the missing intracluster medium is no longer an issue.}


\maketitle{}


\section{Introduction}
\label{sec:intro}

The oldest and most massive star clusters, globular clusters (GC), are expected to host a negligible amount of gas and dust (or intracluster medium, ICM). The notion is so well accepted that the terms ``average'', ``total'', and ``foreground'' reddening have become fully interchangeable when referring to GCs \citep{harris96,harris10,bonatto13}. Globular clusters are indeed expected to lose their natal gas early in their evolution, within $\lesssim$3~Myr according to \citet[][and references therein]{cabrera15}. However, the slow stellar winds of bright red giants, at 10--20~km\,s$^{-1}$ \citep[see e.g.,][]{boyer09,mcdonald19} are expected to accumulate in the GC cores \citep{angeletti82}
in between subsequent Galactic disk crossings. At each disk crossing, the medium is expected to be removed by ram-pressure stripping \citep{chantereau20,naiman20}. Thus in old ($>$10 Gyr) GCs with central escape velocities as high as $\simeq$90 km\,s$^{-1}$ \citep{baumgardt18}, a thin but detectable ICM of about 10$^{-3}-$10$^{-1}$~M$_{\rm{\odot}}$ of dust and 10$^{-3}-$10~M$_{\rm{\odot}}$ of gas is expected \citep{tayler75}, depending on the GC total mass and on when the last disk crossing occurred.

   \begin{figure*}[t]
   \centering
   \includegraphics[width=\textwidth]{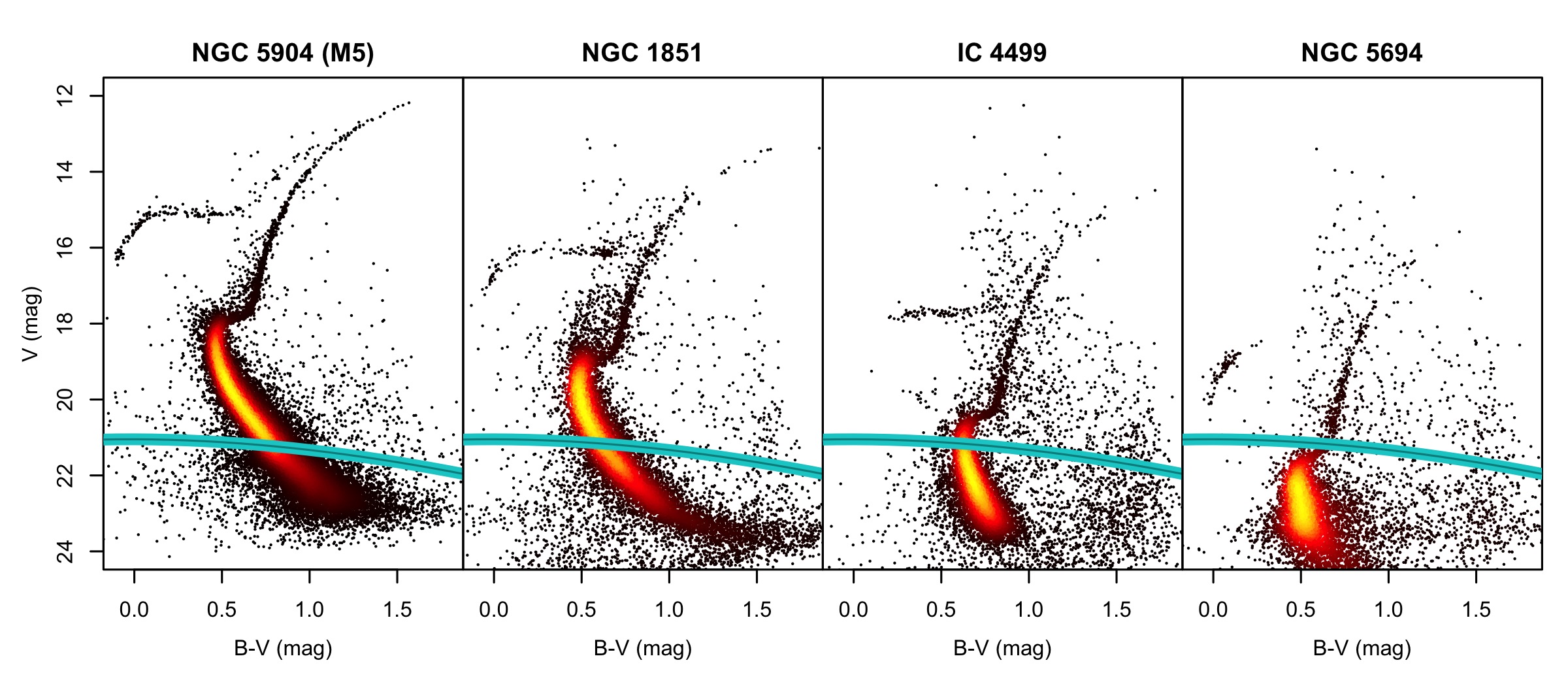}
      \caption{Examples of color-magnitude diagrams from \citet{stetson19}, for GCs at different distances. The {\em Gaia} limiting magnitude (G$\simeq$21~mag) is plotted in all panels as a dark cyan line, with its uncertainty interval in cyan, based on the color transformations by \citet{pancino22}.}
   \label{fig:depth}
   \end{figure*}

Unfortunately, 50 years of observations of GC cores, using a variety of techniques \citep[see, e.g.,][]{knapp96,barmby09}, provided mostly upper limits to the ICM content, which were all a few orders of magnitude smaller than expected on the basis of stellar evolution theory and mass loss among bright red giants. Only a few detections of dust or gas have been claimed so far in NGC\,362 \citep{mendez89}, NGC\,2808 \citep{faulkner91}, NGC\,6624 \citep{grindlay77,forte92}, NGC\,7078 \citep[M\,15,][]{evans03,boyer06}, and NGC\,104 \citep[47\,Tuc,][]{freire01,abbate18}.
In the case of NGC\,362 and NGC\,6624, the photometric detections are really just crude estimates, based on the presence of dark patches and on the color and polarimetry of optical data. For NGC\,2808, the H\,I detection seems secure, but its association with the GC was not fully confirmed. For NGC\,6624 there is also an H$_{\rm{\alpha}}$ detection of ionized gas \citep{grindlay77}. M\,15 is the first GC for which dust was directly detected, through its thermal emission in the infrared with Infrared Space Observatory \citep{evans03}. The dust mass estimate was later revised upward, that is, doubled, by means of Spitzer observations \citep{boyer06}. Finally, in the case of 47\,Tuc, ionized gas was detected with a completely different method, based on the study of its pulsars \citep{freire01}. The estimate was later heavily revised with a larger set of pulsar timing data, thus increasing by almost an order of magnitude \citep{abbate18}. Interestingly, the analysis by \citet{abbate18} highly disfavors models in which the ICM follows the stellar distribution, which was assumed in all of the other cited studies. What is particularly interesting in the literature data is that the detections are either compatible, or in the case of M\,15 only marginally inconsistent, with the mentioned ICM production expectations. Thus the inconsistency between observations and expectations is confined to the upper limits in the literature, while it does not affect the (few) actual detections.

The evolution of stars (both single or in a star cluster) is intimately linked with the medium in which they are embedded, since it regulates the astrochemical processes and the thermal and dynamic state of the system. The admixture of dust and gas affects the overall evolution, and the processes involved are different in the diffuse medium of the Galaxy, GCs, and open clusters. It is therefore crucial to study the properties of the ICM and the gas production and removal mechanisms along the history of a star cluster. For this reason, the discrepancy between the very low upper limits placed on the ICM mass by infrared studies and the theoretically expected masses stimulated a number of theoretical and observational studies. On the one hand, several studies focused on reliably estimating how much gas and dust is actually produced by bright red giants \citep[see][and references therein]{mcdonald19}. On the other hand, theoretical studies on the removal of dust from GCs were carried out \citep{priestley11,chantereau20,naiman20} showing that, unless GC-specific mechanisms are foreseen besides ram-pressure stripping, there should definitely be a thin, but detectable, ICM in most GCs. Possible additional mechanisms were explored by \citet{pepe16}, who proposed that intermediate-mass black holes in the GC can accrete part of the ICM, or by \citet{umbreit08}, who proposed that close stellar interactions could add kinetic energy to the ICM, facilitating its expulsion.

Here we present maps of differential reddening (DR), that is to say variations around the mean reddening, for the 48 GCs studied by \citet{stetson19} and we used them to estimate the mass of the dust contained in the ICM. The paper is organized as follows: in Sect.~\ref{sec:dr} we describe our method for estimating the amount of DR; in Sect.~\ref{sec:maps} we present and validate our DR maps and total DR estimates; in Sect.~\ref{sec:fore} we discuss the nature of the observed DR features; in Sect.~\ref{sec:mass} we estimate the mass of the ICM and compare them with the literature and with theoretical expectations; in Sect.~\ref{sec:disc} we discuss our findings in the framework of the past literature; finally, in Sect.~\ref{sec:concl} we summarize our results and draw our conclusions.


\section{Differential reddening method}
\label{sec:dr}

Methods for DR correction were developed and applied by several authors in the past \citep[for example by][]{turner73,piotto99,vonbraun01,alonsogarcia11,milone12,massari12,bonatto13,dalessandro18}.
The basic idea, common to most methods, is the following: {\em (i)} to define "fiducial" sequences in the color-magnitude diagram (CMD); {\em (ii)} to measure the typical color excesses of stars with respect to those sequences; {\em (iii)} to map how the color excesses (systematically) vary across the GC extent on the sky; and {\em (iv)} to apply the appropriate correction to the magnitudes of each star. Each author developed different implementations of the method, making slightly different choices at each step. We describe our own choices in the following.

\begin{table}
\caption{Cluster sample, with relevant properties from the literature and our new estimated quantities. Our estimates are accompanied by their uncertainties in the electronic version of the table.}
\label{tab:gcs} \label{tab:redd}
\centering                         
\begin{tabular}{lcl}        
\hline\hline                
Column              & Units & Description \\
\hline  
Cluster             &          & GC name \\
AltName             &          & Alternate GC name \\
RA                  & (hrs) & Central RA \citep{stetson19} \\
Dec                 & (deg) & Central Dec \citep{stetson19} \\
X$_0$               & (pix) & Central X \citep{stetson19} \\
Y$_0$               & (pix) & Central Y \citep{stetson19} \\
D                   & (kpc) & Distance$^a$ \\
r$_c$ & ($^{\prime}$) & Core radius$^b$ \\
r$_h$ & ($^{\prime}$) & Half-light radius$^b$ \\
r$_t$ & ($^{\prime}$) & Truncation (tidal) radius$^b$ \\
r$_{\ell}$ & ($^{\prime}$) & Limiting radius (Sect.~\ref{sec:smooth}) \\
\hline
[Fe/H] & (dex) & Metallicity from \citet{harris10} \\
E(B--V)$_{\rm{lit}}$ & (mag) & Average reddening  \citep{harris10} \\
$k$                  & & Number of neighbors (Sect.~\ref{sec:smooth}) \\
dE(B--V)$_{\rm{max}}$ & (mag) & Reddening variation (Sect.~\ref{sec:totdr}) \\
flag & & Foreground contamination \\
& & ("clean", "foreground", \\
& & "high-reddening", see Sect.~\ref{sec:ind}) \\
\hline
n$_{\rm{HB}}$ & & Number of HB stars (Sect.~\ref{sec:theo}) \\
$\tau_{\rm{HB}}$ & (yr) & HB lifetime (Sect.~\ref{sec:theo}) \\
$\tau_c$             & (yr) & Last disk crossing (Sect.~\ref{sec:theo}) \\
th\_n$_{\rm H}^{\rm{min}}$& (cm$^{-2}$) & Expected min N$_{\rm{H}}$ (Sect.~\ref{sec:theo}) \\
th\_n$_{\rm H}^{\rm{max}}$     & (cm$^{-2}$) & Expected max N$_{\rm{H}}$ (Sect.~\ref{sec:theo}) \\
thM$_{\rm{gas}}$     & (M$_{\odot}$) & Expected gas mass (Sect.~\ref{sec:theo}) \\
thM$_{\rm{dust}}$    & (M$_{\odot}$) & Expected dust mass (Sect.~\ref{sec:theo})\\
n$_{\rm H}$       & (cm$^{-2}$) & H column density (Sect.~\ref{sec:dust}) \\
M$_{\rm{gas}}$       & (M$_{\odot}$) & Gas mass estimate (Sect.~\ref{sec:dust}) \\
M$_{\rm{dust}}$      & (M$_{\odot}$) & Dust mass estimate (Sect.~\ref{sec:dust}) \\
\hline 
\multicolumn{3}{l}{$^a$ GC distances from \citep{vasiliev21}.}\\
\multicolumn{3}{l}{$^b$ GC radii from \citet{harris10} and \citet{deboer19}.}\\
\end{tabular}
\end{table}

\subsection{Wide-field photometry}
\label{sec:phot}

We based our DR analysis on the UBVRI Johnson-Kron-Cousins photometry published for 48 globular clusters (GC) by \citet{stetson19}. The list of GCs is presented in Tab.~\ref{tab:gcs}, along with adopted quantities from the literature and new quantities computed here. The catalogs are based on about 90\,000 public and proprietary CCD images, which are extracted from a larger database of about one million, collected to create an accurate set of secondary standards in the Johnson-Kron-Cousins photometric system \citep{stetson00,pancino22}\footnote{\url{https://www.canfar.net/storage/list/STETSON}}. The strengths of the photometric catalogs used here are the wide area covered (35--80'); the photometry accuracy (0.005--0.020~mag, depending on the passband); and the depth of the photometry (V$<$23--25~mag). The photometry is ground-based, so the central regions suffer from stellar crowding, with the typical seeing being $\simeq$1". 

We performed a relaxed quality selection based on the following criteria \citep[see][for details]{daophot1,stetson88}: {\em (i)} BVI magnitude uncertainties smaller than 0.5~mag; {\em (ii)} $\chi<5$; and {\em (iii)} image sharpness parameter $|$sharp$|<1$. This resulted in the removal of about 10--40\% of the sources from each catalog, with preferential removal of the faintest ones. We aim at a homogeneous analysis, thus we focused on the BVI bands, because the U and R bands are often not as deep, not allowing a good sampling of the main sequence (MS) in several GCs. An example of the CMDs of four selected GCs, at different distances, is presented in Fig.~\ref{fig:depth}, where the {\em Gaia} magnitude limit is also reported, which is too shallow for our goals (see below).

   \begin{figure}[t]
   \centering
   \includegraphics[bb=0 50 2800 3300,width=0.8\columnwidth]{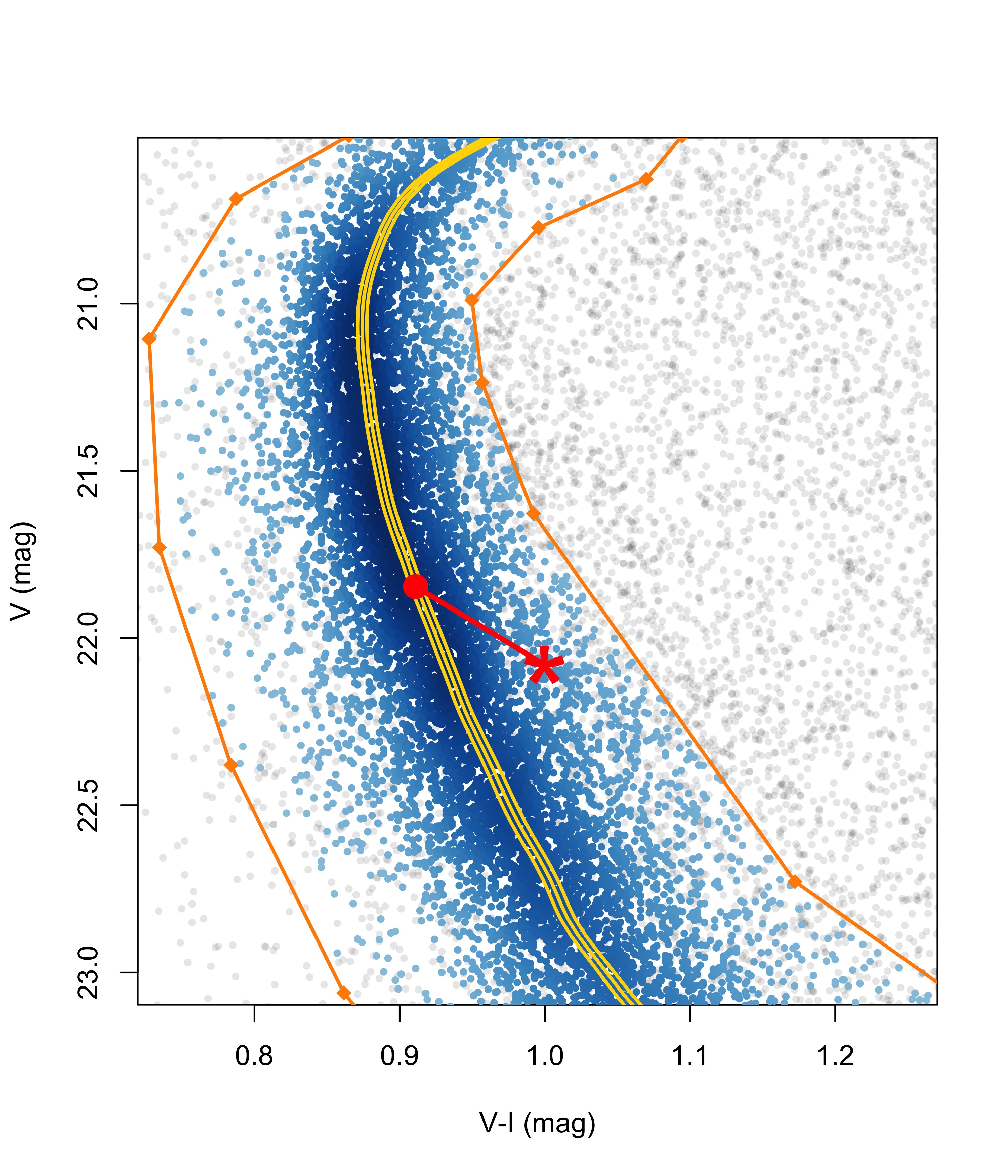}
   \vspace{-0.3cm}
      \caption{Illustration of the DR method, using the case of IC\,4499 in the (V,V--I) CMD plane as an example. Gray dots show the full reference photometric catalog. The manual preselection box is plotted in orange. The sample of stars belonging simultaneously to the four boxes in the (V,B--V), (V,V--I), (V,B--I), and (B--V,V--I) planes is plotted in blue. The median ridge line (MRL) is plotted as a yellow curve, bracketed by its uncertainty curves. The color displacement $\Delta$E(V--I), is computed for each star (red star symbol) along the reddening line (red line), which intersects the MS on the red dot. This is the position  that the star would have on the MRL if no reddening or photometric errors were present.}
   \label{fig:box}
   \end{figure}

\subsection{Main sequence samples}
\label{sec:ms}

The photometry is deep enough to cover from 2 to 5 magnitudes below the Turn-Off (TO) point, with a median count of $\simeq$10\,000 stars per GC (see Fig.~\ref{fig:depth} and Tab.~\ref{tab:maps}). 
In the past, the less populated red giant branch or horizontal branch were successfully used when necessary \citep{melbourne00,gerashchenko04,contreras13}, but whenever the MS is well observed, it certainly provides a far richer sample, covering much better the GC extent. 

We manually defined rough preselection boxes around the MS in the (V,B--V), (V,V--I), and (V,B--I) CMD planes and in the (B--V,V--I) color-color plane. Stars lying within all four boxes defined our ``MS samples'' (Fig~\ref{fig:box}). The simultaneous multicolor selection allowed us to remove most of the field stars, whose sequences have different color offsets from the MS of GCs in the four planes, because of the different metallicity, age, or evolutionary status. For most of the GCs we could also efficiently separate and remove the binary sequence, which is the sequence of stars redder and brighter (by 0.75~mag), running parallel to the MS and composed by the blended images of equal-mass binaries (see also Sect.~\ref{sec:spurious}). 

We could not use {\em Gaia} proper motions to select bona-fide members, because they are not available for stars fainter than V$\simeq$21~mag (Fig.~\ref{fig:depth}) and thus we would have to drop 34 of our 48 GCs and use much less populous samples. Recently, the \citet{stetson19} sample was analyzed by \citet{jang22}, using {\em Gaia} selected members, to determine DR maps for a different scientific goal. To avoid dropping too many distant GCs, they included red giants and subgiants, so they were only forced to exclude five sparsely populated GCs. However, this means that their DR estimates for nearby GCs are dominated by MS stars and the ones for distant ones by red giants, while we need the maximum possible homogeneity in the analysis to fulfil our scientific goal. Additionally, our MS samples comprise generally more stars, so our DR maps have typically higher spatial resolution, which is needed for estimating ICM masses (Sect.~\ref{sec:dust}).

To test our bona fide members selection procedure, we used {\em Gaia} EDR3 data for the 14 GCs with the TO brigther than the {\em Gaia} limit by at least 2~mags. We noticed that our lists of photometric members overlap the ones by \citet{vasiliev21} by at least 75\%, depending on the distance. For example, in the case of the closest GC, NGC\,6121 (M\,4), our starting MS sample contains 16\,781 stars above V=21~mag, of which 15\,805 (i.e., 94\%) have $>$80\% membership probability by \citet{vasiliev21}. This confirms that our photometric sample selections are as clean as possible from field star contaminants.

\subsection{Median ridge lines}
\label{sec:mrl}

Different recipes were successfully used in the past to compute fiducial sequences, including the use of theoretical models. We chose to use empirical sequences \citep[see also][]{milone12}, given that our MS are well populated and defined. Because the color distribution around the MS is asymmetric and heteroscedastic, we computed median ridge lines (MRL) rather than mean ones. For the sole purpose of computing the MRL, we cleaned our MS samples using more stringent criteria compared to Sect.~\ref{sec:ms}: {\em (i)} uncertainties on BVI smaller than 0.3~mag, {\em (ii)} $\chi\,<\,$3, and {\em (iii)} $|$sharp$|<$0.5. We also excluded stars closer to the GC center than 0.5\,r$_c$ (core radii) or farther than 0.8\,r$_t$ (truncation radii, see Tab.~\ref{tab:gcs}). This way, the crowded core and the outer field-contaminated parts of the GC were not used to compute the MRL. For the most crowded or reddened GCs it was necessary to adopt stricter radial selections: in those cases we used the behavior of V and $\sigma_{\rm{V}}$ as a function of radial distance to identify the radius at which crowding starts to be important.

The MRL was computed in each of the three (V,B--V), (V,V--I), and (V,B--I) planes. We first computed a running median along V magnitude, using bins of about 20--1000 stars, depending on how populous was the MS sample of each GC. We then fit a smoothed cubic spline on the weighted median points, where the optimal smoothing parameter $\lambda$ was found using a generalized cross-validation algorithm. We iteratively applied a 1--99 quantile clipping, rejecting stars that were outside the quantile range in at least two colors. Statistical uncertainties on the MRL were computed by fitting a smooth spline
to the median absolute deviation of the V bins and are typically on the order of 0.01~mag or less. Systematic uncertainties were estimated by fitting a smooth spline to MRL$_{\rm{BV}}$+MRL$_{\rm{VI}}$-MRL$_{\rm{BI}}$ and are typically between 0.001 and 0.01~mag. An example of MRL computation for IC\,4499 in one of the planes is shown in Fig.~\ref{fig:box}.

\subsection{Raw color displacements}
\label{sec:raw}

We computed raw color displacements\footnote{Color displacements, as well as reddening variations, can be negative because they are computed with respect to the MRL or the mean E(B--V) of the GC, respectively.} of each star in each GC from their MRLs, using the whole starting MS sample (Sect.~\ref{sec:ms}). We did so in each of the three CMD planes separately: (V,V--I), (V,B--V), (V,B--I). The displacements were obtained as the difference in color between each star and the point that the star would occupy if it were not reddened and if there were no photometric errors (Fig.~\ref{fig:box}). Such point was obtained as the intersection between the MRL and the reddening line passing through each star. 

We assumed the classical R$_V$ value for the reddening line: R$_V$=A$_V$/E(B--V)=3.1, E(V--I)=1.244\,E(B--V), and E(B--I)=2.244\,E(B--V) \citep{dean78}. These classical relations are only valid on average, given the variety of the interstellar medium in the Galaxy. Therefore, to test the choice, we recomputed our DR maps for half a dozen GCs, using the R$_V$ and color relationships appropriate for the line of sight of M\,4, which lies behind the Sco-Oph dark cloud \citep{hendricks12} and thus its R$_V$ is about 10\% higher. As a result, the final DR values in our maps changed only on the fourth or fifth decimal digit.

A raw DR estimate, $\Delta$E(B--V), was obtained as the  weighted average of the estimates in the three CMD planes, reported to E(B--V) using the classical relations above. We used the photometric and the MRL errors for the weights computation. We did this to average out as much as possible spurious features (see Sect.~\ref{sec:maps}). We aim in fact at the maximum accuracy, rather than at the maximum precision \citep[unlike][]{milone12,massari12,dalessandro18}. 
Our typical MS samples contain dE(B--V) estimates for about 10\,000 MS stars (median), with the least sampled GC being Pal\,14 ($\simeq$850 stars) and the most sampled NGC\,5139 ($\simeq$156\,000 stars). 

\subsection{Spatial smoothing}
\label{sec:smooth}

Photometric errors, with the exception of the crowded GC centers (see Sect.~\ref{sec:spurious}), do not vary significantly with sky coordinates within the GC. The reddening caused by the interstellar medium, on the other hand, changes with position if the medium is not distributed homogeneously. A smoothing in sky coordinates is thus an effective way to make the spatial reddening variations emerge from the noise of photometric uncertainties. Among the methods employed in the literature, we chose to apply a median-smoothing technique, similar to the one adopted by \citet{milone12}, \citet{massari12}, or \citet{dalessandro18}. This choice is motivated by the need to preserve as much as possible the higher spatial resolution in the densest parts of the GC. 

The raw $\Delta$E(B--V) obtained for each star was thus replaced with the median value of its $k$ nearest neighbors, dE(B--V). The choice of $k$ was iteratively adjusted for several GCs, to simultaneously minimize the noise in the DR maps and to maximize the tightening of the CMD sequences. The final $k$ values, mostly around 200, are reported in Tab.~\ref{tab:redd} for each GC. This way, it was also possible to compute -- in each star's neighborhood -- the median absolute deviation $\sigma$E(B--V), the formal statistical error $\delta$E(B--V)=$\sigma$E(B--V)/$\sqrt(k)$, and the local resolution $\rho$E(B--V), determined as half the distance to the most distant neighbor\footnote{Interestingly, we note that the map errors, $\sigma$E(B--V), are significantly higher for stars whose neighborhood includes sharp reddening variations, and can thus be used as edge detectors in the maps if needed.}.
When computing the maps, we also computed a "limiting radius" (r$_{\ell}$, Tab.~\ref{tab:gcs}), at which the number of stars in the GC roughly equals the number of field stars. At r$_\ell$ or beyond, the DR maps are not reliable anymore and we flagged those stars with {\tt flag\,=\,-1} (see Tab.~\ref{tab:maps}). However, we suggest to cut the map earlier to be on the safe side, for example at 80\% of r$_{\ell}$.

\begin{table}
\caption{Differential reddening maps. For each star in the MS samples we provide coordinates and V magnitudes from \citet{stetson19}, and other quantities derived as described in Sect.~\ref{sec:dr}. The DR map itself can be obtained considering only the star coordinates and dE(B-V).}
\label{tab:maps} 
\centering                         
\begin{tabular}{lcl}        
\hline\hline                
Column              & Units & Description \\
\hline  
Cluster             &          & GC name \\
Star                &          & Star ID, unique in each GC \\
X                   & (")      & Pixel coordinate along RA \\
Y                   & (")      & Pixel coordinate along Dec \\
RA                  & (deg)    & Right Ascension in decimal degrees \\
Dec                 & (deg)    & Declination in decimal degrees \\
V                   & (mag)    & V-band magnitude of the star \\
$\Delta$E(B--V)     & (mag)    & Raw color difference \\
dE(B--V)            & (mag)    & Median-smoothed DR \\
$\sigma$E(B--V)     & (mag)    & Median absolute deviation on dE(B--V) \\
$\delta$E(B--V)     & (mag)    & Statistical error, or $\sigma$E(B--V)/$\sqrt(k)$ \\
$\rho$E(B--V)       & (")    & Local resolution of dE(B--V) \\
flag & & Quality flag (--1\,=\,outside r$_{\rm{o}}$; 0\,=\,good; \\
\multicolumn{3}{r}{1\,=\,instrumental; 2\,=\,stellar)} \\
\hline                                    \end{tabular}
\end{table}

   \begin{figure*}[t]
   \centering
   \includegraphics[width=\textwidth]{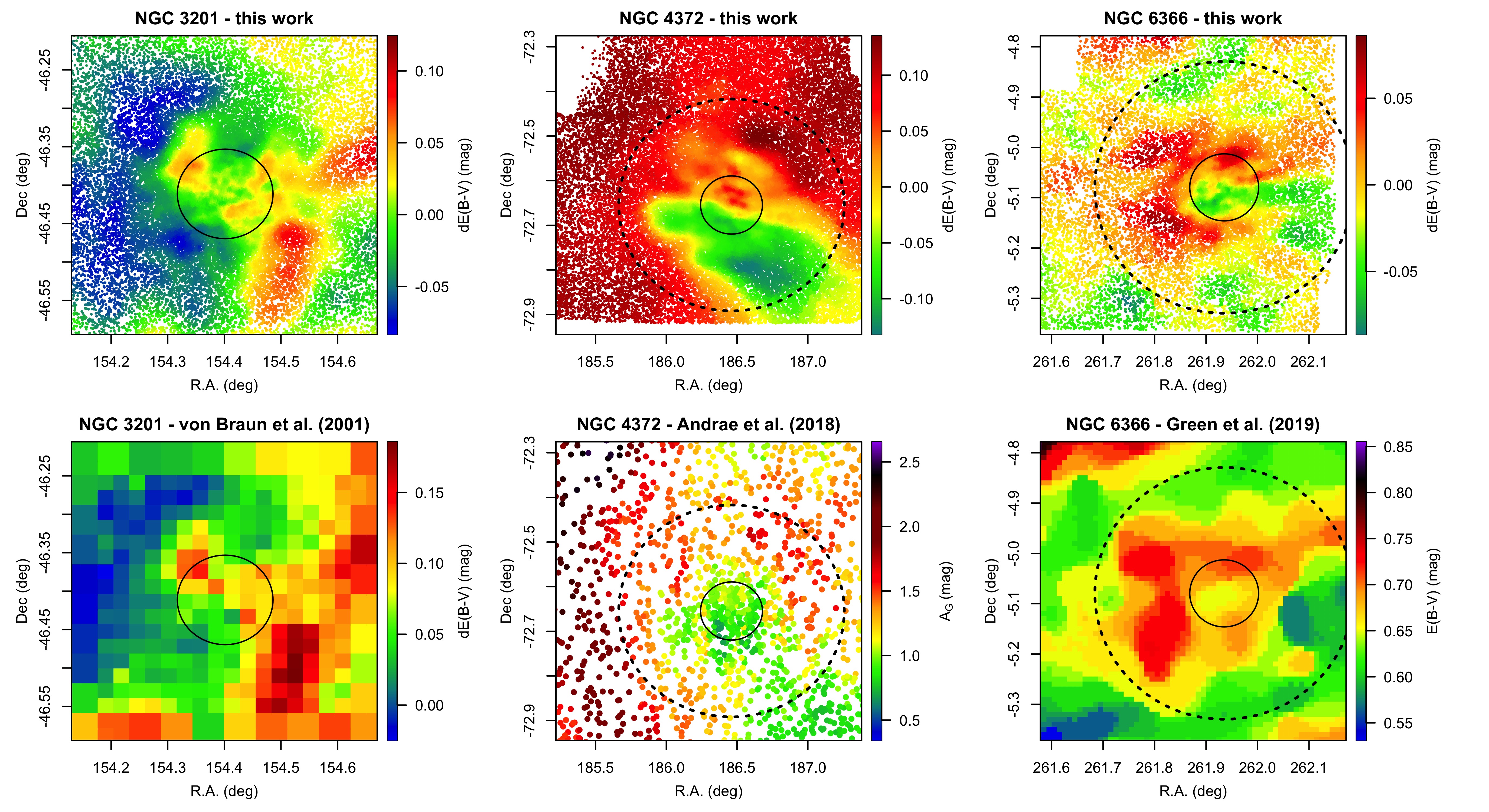}
      \caption{Comparison of our DR maps (top row) with literature ones (bottom row) for three well-studied GCs, as annotated. The left panels show NGC\,3201, compared with the optical map obtained with a method similar to ours \citep{vonbraun01}. The center panels show NGC\,4372 compared with the Gaia DR2 A$_{\rm{G}}$ values for stars brigther than G$\simeq$17 mag \citep{andrae18}. The right panels show NGC\,6366 compared with the 3D reddening maps obtained from Pan-STARSS, 2MASS, and Gaia data \citep{green19}. The coordinate and color scales are tailored to match the different literature sources. In all panels, the solid circle represents r$_h$ and the dashed circle – when visible – r$_{\ell}$.}
   \label{fig:lit1}
   \end{figure*}

   \begin{figure}[t]
   \centering
   \includegraphics[width=\columnwidth]{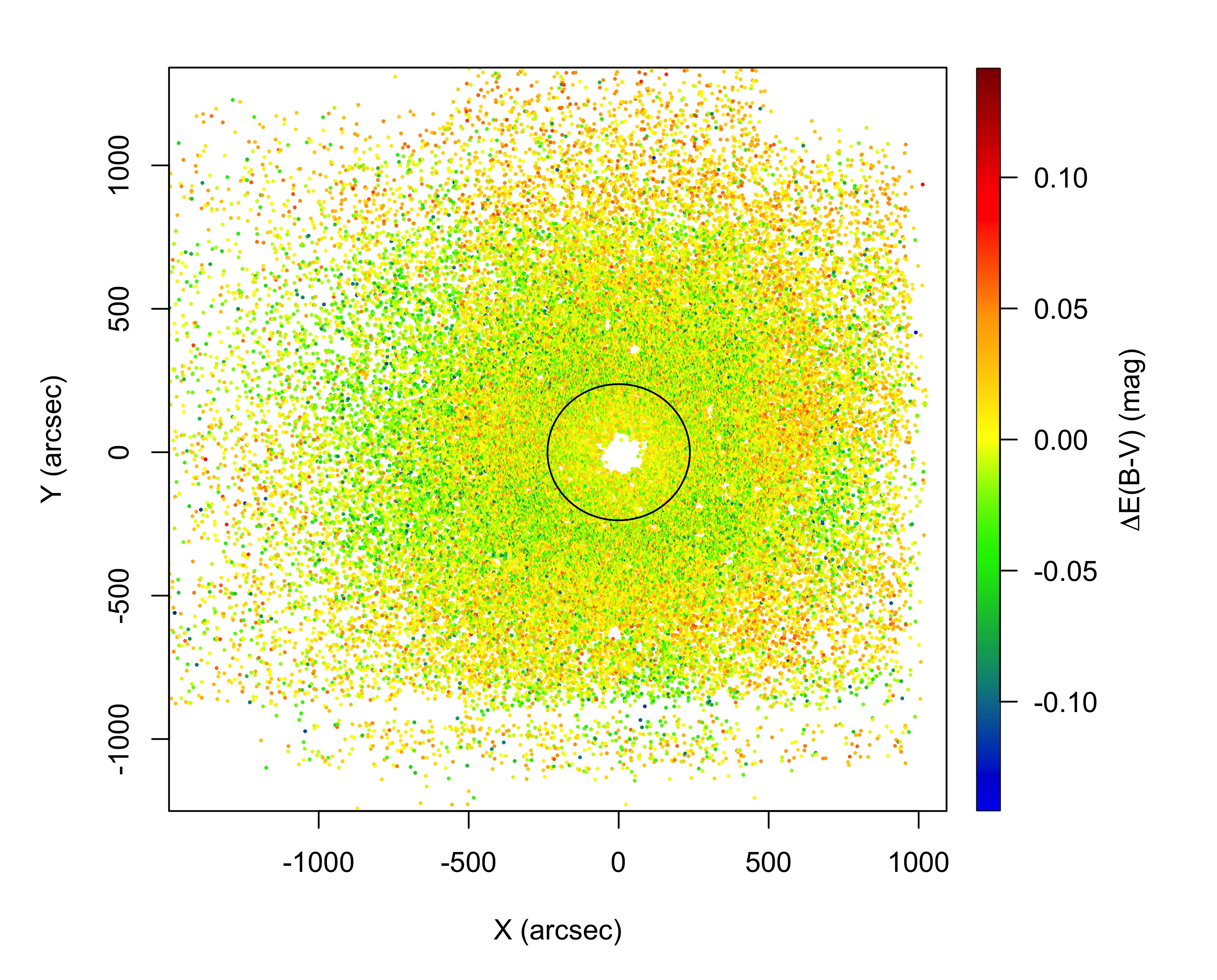}
   \vspace{-0.5cm}
      \caption{Example of (raw) DR map dominated by instrumental features, in the case of NGC\,104 (47\,Tuc). The footprint of the CCD frames included in the original wide field photometry \citep{stetson19} is clearly visible. The half-light radius (r$_h$) is plotted for reference. The hole in the GC center is caused by stellar crowding.}
   \label{fig:spurious}
   \end{figure}


\section{Differential reddening results}
\label{sec:maps}

As a result of the procedure outlined in Sect.~\ref{sec:dr}, we obtained the DR maps, presented in Tab.~\ref{tab:maps} and plotted in Figs.~\ref{fig:allmaps1}--\ref{fig:allmaps4}, which we carefully compared with publicly available maps from various sources cited throughout this paper. Unfortunately, the maps by \citet{jang22}, which would be an ideal comparison set because they are also obtained from \citet{stetson19} photometry, are not publicly available. However, by visually comparing their Figs.~4 and 5 with our maps, we note that they are nearly identical, as expected given the common dataset and similar method. Some examples of literature comparisons are shown in Fig.~\ref{fig:lit1}. As can be seen, our maps are qualitatively very similar to the literature ones, regardless of the exact method employed, but in general they tend to have higher spatial resolution. 

The zero point (ZP) of our maps is set by the choice to compute color displacements from the MRL, thus it corresponds more closely to the median E(B--V) reddening of the GC, which does not always correspond exactly to the average E(B--V) available, for example, in the \citet{harris10} catalog (see Tab.~\ref{tab:gcs}). The largest differences, if any, are expected for heavily reddened GCs, especially when the DR distribution is far from being Gaussian.


\subsection{Spatial ZP variations}
\label{sec:zp}

As discussed by \citet[][see their Section 3.2]{stetson19}, wide field imagers can suffer from various instrumental effects, including light concentration, that ultimately cause a ZP variation from the margin of the field of view to the central parts. To minimize these effects, the original multiband photometric catalogs obtained from individual CCD images were corrected for spatial ZP variations with bilinear functions, but for some imagers this left residual ZP variations, whose footprint is well visible in some of our reddening maps. 

One such case is illustrated in Fig.~\ref{fig:spurious}, where the rectangular footprint of the CCDs is visible in the non-smoothed DR map of 47\,Tuc. The amplitude of the patterns ranges roughly from 0.005 to 0.020~mags (i.e., they are comparable to the photometric uncertainties) and thus they can only be noticed for GCs with little DR. This of course does not exclude that similar spurious patterns are also hidden in the reddening maps of GCs with much larger reddening variations, but in that case they become comparatively irrelevant. Noteworthy, there are GCs with very small reddening variations that do not show any spurious rectangular pattern corresponding to any of the CCD footprints. We manually flagged stars in the reddening maps across these spurious patterns with {\tt flag\,=\,1} in Tab.~\ref{tab:maps}. When applying the maps to the photometry by \citet{stetson19} these stars could be included, with the added benefit to mitigate the ZP variations. Otherwise, if the maps are applied to different photometric catalogs or used for other purposes, these stars should not be considered.

In particular, in the cases of NGC\,5024 and NGC\,6809, we observe suspiciously squared DR patterns, but since we could not link them to any of the known CCD footprints in the relevant datasets, we did not flag any star. On the other extreme, NGC\,104 (47\,Tuc) and NGC\,6752 show evident DR patterns that are readily connected with some of the CCD footprints in their datasets, so we flagged their stars. Another two GCs, NGC\,5139  ($\omega$\,Cen) and NGC\,5272, show moderate patterns, but still connected with some of the CCD footprints, which were also flagged. In all cases, the CCD footprint and spurious DR variations are outside of the GC half-light radii and therefore have no impact on our ICM mass determinations (Sect.~\ref{sec:dust}).

   \begin{figure}[t]
   \centering
   \includegraphics[width=\columnwidth]{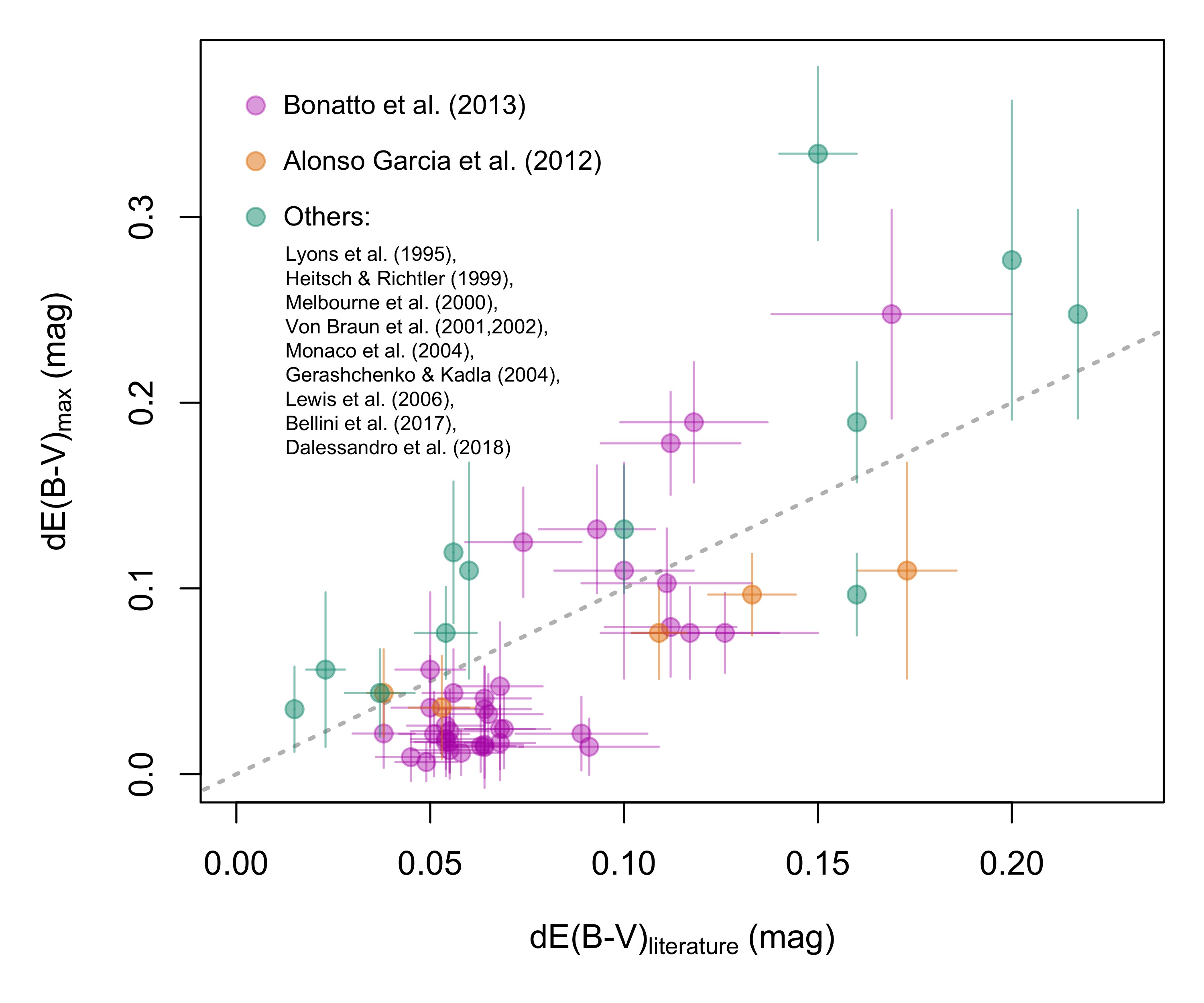}
   \vspace{-0.5cm}
      \caption{Comparison of the total reddening variations, dE(B–V)$_{\rm{max}}$, found here and in the literature collection. Different colors refer to different sources, as labeled. The 1:1 relation is indicated as a dotted gray line. See Sect.~\ref{sec:totdr} for more details.}
   \label{fig:lit2}
   \end{figure}

\subsection{Stellar population effects}
\label{sec:spurious}

Besides reddening and photometric uncertainties, there are astrophysical reasons why a star can be displaced from the MRL. In the particular case of GCs, we can list three main stellar population effects: {\em (1)} the sequence of unresolved equal-mass binaries, which appears brighter than the MS by 0.75~mag, and thus are redder at a given magnitude; {\em (2)} the effect of stellar crowding, which increases toward the GC center; when considering that upper MS stars are bluer than the rest of the sequences, there is a higher chance for them to be blended with redder stars; and {\em (3)} last but not least, the effect of multiple stellar populations in GCs, which produce multiple sequences at different colors across the entire CMD \citep{bedin04,milone12,bellini17b}.

The problem of multiple populations in GCs has been extensively reviewed \citep{gratton12,bastian18}. What is of relevance here, is that the different chemical composition of stars belonging to different populations in the same GC, especially as far as He and CNO elements are concerned, generates multiple photometric sequences \citep{pancino00,sbordone11,milone17}. The typical separation in optical colors of the multiple MS is on the order of $\lesssim$0.1~mag, and it is higher only in the case of exceptional GCs such as, for example, NGC\,2808. It has been shown that the enriched populations can be more centrally concentrated \citep{lardo11,leitinger23} with few exceptions, so if they play a role, they are expected to produce a round central region of different reddening. For binaries, we could easily separate the equal-mass binary sequence from our starting sample in most GCs, but in a few cases of heavily reddened and contaminated GCs this was not so easy to achieve. We expect unresolved binaries and stellar crowding effects to produce a round DR enhancement at the GC center, where binaries are concentrated because they are more massive than single stars and crowding effects are more severe. 

Unfortunately, there is no easy way to separate the cases in which a central region with higher reddening is caused by binaries, crowding, or multiple stellar populations from the cases in which it is genuinely caused by DR. Therefore, we identified four GCs in the sample with round (i.e., radially symmetric) concentrations of high DR compared to the surrounding region. Of these, none presented problems in separating the equal-mass binary sequence from our MS samples, but the size of the central reddened regions correspond to the radius at which, in the (X,V) plane, there is a sudden incompletess toward the center. The GCs are IC\,4499, 47\,Tuc, $\omega$\,Cen, and M\,22, of which the last three are known to display exceptional multiple populations. For these reasons, we marked the central regions of these GCs with {\tt flag\,=\,2} in Tab.~\ref{tab:maps} and in some cases it was not possible to compute a reliable dust mass estimate (Sect.~\ref{sec:dust}). In any case, it is always advisable to be extra cautious with ground-based photometry of the central few arcmin of GCs, including our own DR maps.

\subsection{Total reddening variation}
\label{sec:totdr}

To compute the total reddening variation suffered by a GC, dE(B--V)$_{\rm{max}}$, we employed only stars with {\tt flag\,=\,0} (Sect.~\ref{sec:dr}) in our DR maps (Tab.~\ref{tab:maps}), thus excluding regions affected by instrumental or stellar features. To make sure that we used regions dominated by GC stars rather than the field, we further selected stars contained within 80\% of r$_{\ell}$, the radius at which the number of cluster stars equals the number of field stars (Sect.~\ref{sec:smooth}). We estimated the total reddening variation as 5\,$\sigma$ of the DR distribution of the selected stars and the uncertainty was computed as a simple error propagation. The results can be found in Tab.~\ref{tab:gcs}.

We compared our estimates with a collection of literature ones obtained with similar methods, as reported in Fig.~\ref{fig:lit2}. We relied mostly on the work by \citet{alonsogarcia12} and \citet{bonatto13}, who provided results for several GCs, but we also included estimates for individual GCs by \citet{lyons95}, \citet{heitsch99}, \citet{melbourne00}, \citet{vonbraun01,vonbraun02}, \citet{monaco04}, \citet{gerashchenko04}, \citet{lewis06}, \citet{bellini17a}, and \citet{dalessandro18}. We note that the GCs with low dE(B--V)$_{\rm{max}}$ tend to have consistently higher values in \citet{bonatto13} than here, by about 0.03~mag. This agrees well with their estimate of the residual ZP variations in their original photometry, which are apparently higher than the ZP variations in the \citet{stetson19}
photometry used here. Once this is corrected, our estimates generally agree with the literature with a mean difference of 0.002~mag and a 1\,$\sigma$ spread of $\simeq$0.03~mag, which is comparable to our typical random uncertainties. We conclude that the comparison is satisfactory.

\subsection{CMD correction}
\label{sec:cmd}

   \begin{figure}[t]
   \centering
   \includegraphics[width=0.9\columnwidth]{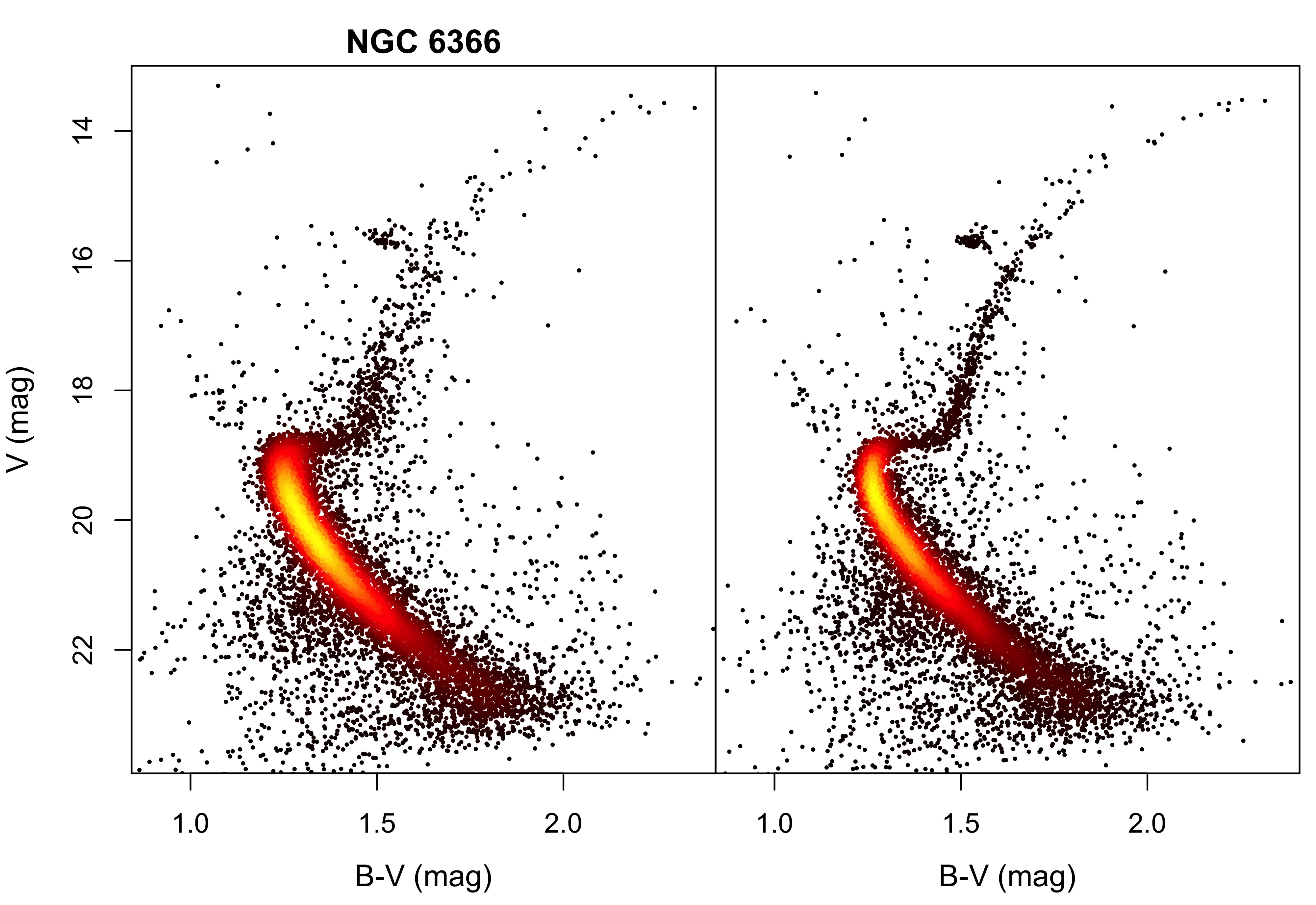}
   \includegraphics[width=0.9\columnwidth]{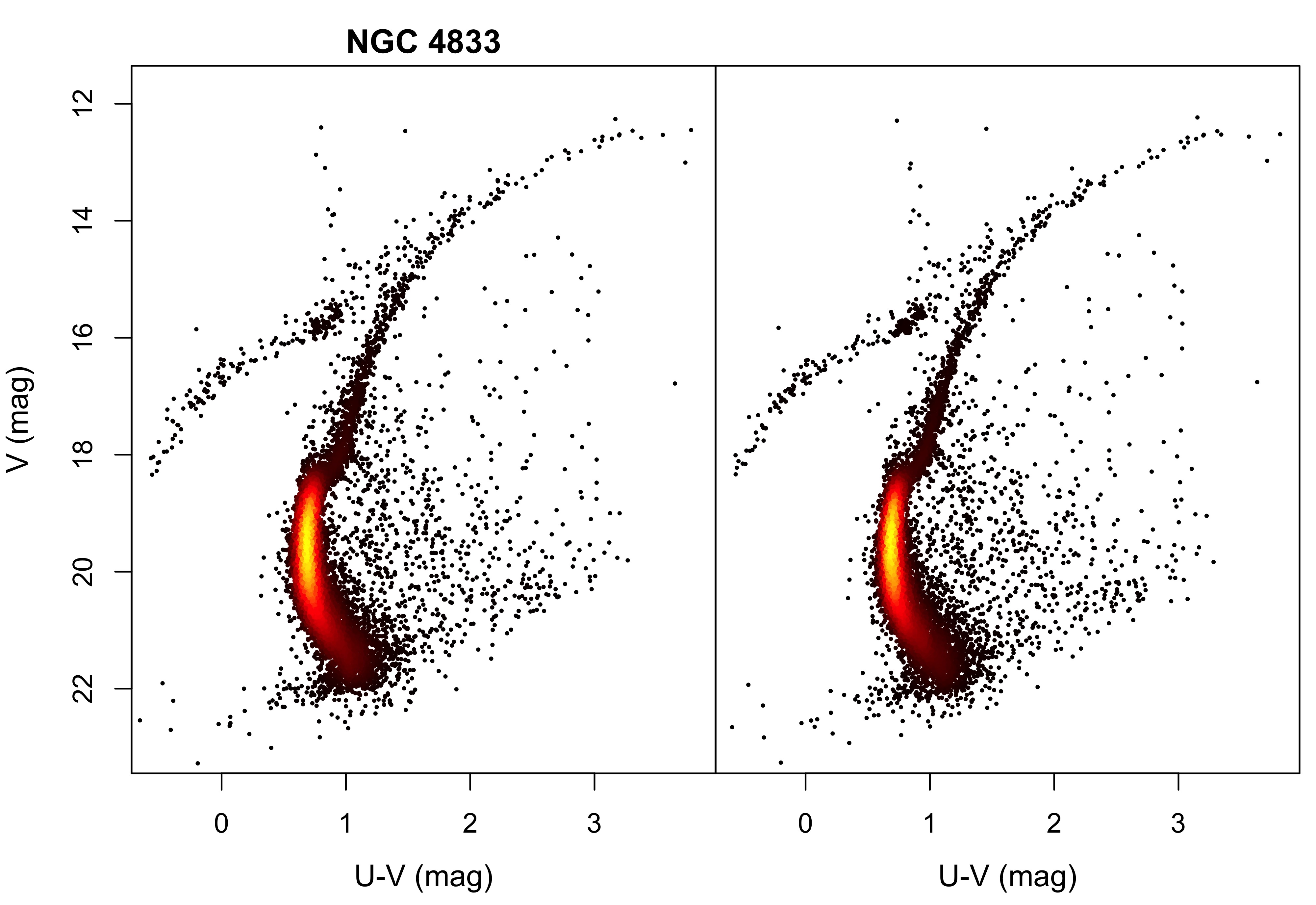}
      \caption{Example of DR correction. The left panels show the original photometry and the right panels the one corrected for differential reddening. The top panels illustrate the case of NGC 6366 in the (V,B–V) CMD plane, the bottom panels show NGC 4833 in the (V,U–V) CMD plane. The points are colored according to the density of stars in the CMD, with black being the lowest density and yellow the highest one.}
   \label{fig:cmdcorr}
   \end{figure}

The most common application of DR maps is their use to make the evolutionary sequences in CMDs appear tighter by removing star-to-star reddening variations. We show an example in Fig.~\ref{fig:cmdcorr}. We summarize here below our recommendations when using our DR maps for this purpose.

\begin{itemize}
    \item{Stars with {\tt flag\,=\,2} or {\tt -1} should be excluded, while stars with {\tt flag\,=\,1} should be included only when correcting the photometry by \citet{stetson19}. As a more restrictive criterion, stars outside 80\% of r$_{\ell}$ could be discarded. }
    \item{Each star that needs correction can be assigned the dE(B--V) value of the closest star (in coordinates) in our DR maps; alternatively, a 2D interpolation of our DR maps could be used to compute the dE(B--V) value at the desired position.}
    \item{For GCs with a total reddening variation (see Sect.~\ref{sec:totdr} and Tab.~\ref{tab:gcs}) comparable to the typical photometric errors for that GC, or smaller, any apparent improvement or worsening of the sequences tightness might not be always statistically significant; the improvement may appear clear in some CMD color planes and counterproductive in other planes; in particular, we advise to excercise caution in correcting GCs with dE(B--V)$_{\rm{max}}\lesssim$0.03~mag, depending on the typical photometric errors for that GC and for the specific evolutionary sequence of interest.}
    \item{Additionally, for GCs with dE(B--V)$_{\rm{max}}$$\lesssim$0.02~mag, our maps could be subject to extra uncertainties related to ZP variations in the \citet{stetson19} photometry (Sect.~\ref{sec:spurious}), at least in some specific colors.}
    \item{For some GCs with large field contamination, such as bulge GCs, the improvement in the evolutionary sequences tightness, although clearly noticeable, might not be completely satisfactory (see also Sect.~\ref{sec:ms}). An example is NGC\,6760. Similarly, it is difficult to evaluate the goodness of the correction for very complex GCs such as $\omega$~Cen or NGC\,2808.}
\end{itemize}


\section{Reddening maps interpretation}
\label{sec:fore}

The morphology of the DR maps (Figs.~\ref{fig:allmaps1}--\ref{fig:allmaps4}) is complex. There are GCs which appear split into two parts with very different reddening, others with stripes and lanes passing very close to the GC center, others showing arcs of higher reddening, which are reminiscent of projected torus-like structures or bubble-like structures. Only in very few GCs we observe the expected central concentration of high reddening \citep{angeletti82}. This suggests that, if the DR is caused by the ICM, the dust is not spherically distributed and in general it is not concentrated toward the GC center, which could be one of the reasons why infrared searches found only low upper limits (see also Sect.~\ref{sec:spat}). We recall at this point that the pulsar data by \citet{abbate18} highly disfavor a concentrated ICM distribution. However, the fundamental question is whether the observed DR structures originate from dust within GCs, or in front of them, or both. Fortunately, we can use 3D reddening maps for the purpose \citep{green19,lallement19,lallement22}, as described in the following.

   \begin{figure}[t]
   \centering
   \includegraphics[width=0.9\columnwidth]{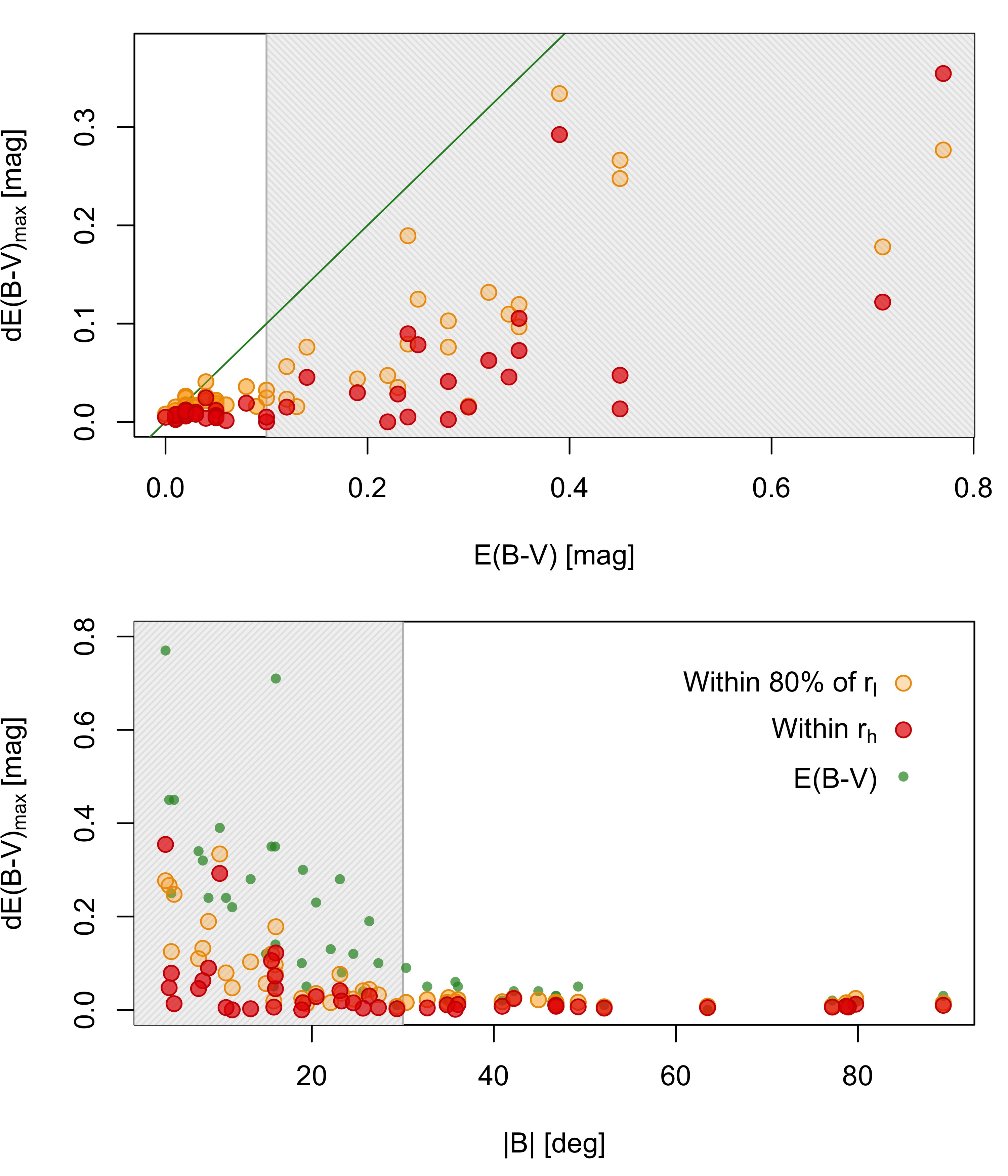}
      \caption{Behavior of the total DR, dE(B--V)$_{\rm{max}}$, with E(B--V) (top panel) and Galactic latitude (bottom panel). In each panel, orange symbols are the total DR computed within 80\% of r$_{\ell}$, red ones within r$_h$. The green line in the top panel is the 1:1 agreement, while the green dots in the bottom panel are the E(B--V) values of the sampled GCs. The gray-shaded areas cover the high-reddening sample (Sect~\ref{sec:circ}).}
   \label{fig:circ}
   \end{figure}

\subsection{Galactic cirri}
\label{sec:cirrus}

If the reddening variations are mostly caused by a foreground component, we argue that this should be the diffuse component and in particular Galactic cirri \citep{low84,magnani85,andre10,heyer15,bianchi17}, because translucent clouds or star-forming filaments would imply a higher extinction and extinction variation than observed in our maps \citep[A$_V$$\gtrsim$2~mag,][]{vandishoeck89}. The highest reddening is observed for NGC\,6760 and it is E(B--V)\,=\,0.77\,mag, while the highest DR is dE(B--V)$_{\rm{max}}$$\lesssim$0.3~mag. Galactic cirri are visible at all latitudes, they contain gas (atomic and molecular) and dust, and are mostly located within a few hundred parsecs from the Sun \citep{devries87,zucker19}. To understand whether it is plausible that all GCs in our sample are contaminated by Galactic cirri, we need to assess the size and density of cirri on the plane of the sky. 

The Galactic medium is made of filaments with widths typically on the order of 0.1~pc \citep{pineda23}. However, if the filaments belong to the foreground medium, what really counts is their projected width: for a typical distance of 100--200~pc, it would be on the order of 3--4~arcmin, which is compatible with some of the structures visible in Figs.~\ref{fig:allmaps1}--\ref{fig:allmaps4}, but not with all of them. However, the closeness of the filaments to the observer makes their apparent sizes vary a lot with distance, so we cannot draw conclusions from the apparent sizes alone.

Using the SDSS Stripe~82 data, covering an area of $\simeq$5\,deg$^2$, \citet{smirnov23} studied Galactic cirri in reflected optical light and found that about 2\% of the surveyed area is covered by cirri, quite unequally distributed, each with a typical (median) area on the order of 1~arcmin$^2$. The density of filaments per square degree changes significantly from region to region, from a minimum of about zero to a maximum of about 250. Because we find substructures within the half-light radii \citep[typically 4~arcmin,][]{deboer19} of each of our 48 GCs, and because the sky coverage of cirri appears so uneven, we find it unlikely that all of the observed substructures in our GC sample are caused by the local foreground medium of our galaxy. We discuss foreground contamination for our GCs in more details in Sect.~\ref{sec:ind}.  

\subsection{Circumstantial evidence}
\label{sec:circ}

We examine in Fig.~\ref{fig:circ} the behavior of the total DR as a function of E(B--V) and Galactic latitude. We observe that low-reddening GCs, as well as high-latitude ones, show a flat trend with both parameters. If the reddening variations of these GCs were entirely caused by foreground dust, we would expect them to correlate with the total amount of reddening. We tested the correlation using the Pearson's, Kendall, and Spearman methods and found that the data are not correlated: the results are in the range --0.32 to --0.48 in the case of Galactic latitude and between 0.31 and 0.43 for E(B--V). High-reddening GCs show, on the other hand, a marked increase of the total DR with increasing E(B--V) or decreasing Galactic latitude. The lack of correlation for the low-reddening or high-latitude GCs becomes even more pronounced if we compute the total DR on a smaller area within each GC, such as within r$_h$ instead of 80\% of r$_{\ell}$ (Sect.~\ref{sec:totdr}). Most importantly, we observe a systematic increase in the total reddening variation if the area covered within each GC increases. If the DR was caused entirely by the foreground medium, such as for example a filament passing in front of the GC, we would not expect any systematic variation of the total DR with the area covered within each GC. This behavior is observed in all the sample GCs, as is apparent from Fig.~\ref{fig:circ}. 

In conclusion, the DR observed in the low-reddening or high-latitude GCs appears not to vary with the amount of total reddening, suggesting that their DR could in principle be a good tracer of their ICM distribution. We thus split our sample at E(B--V)$<$0.1~mag and $|$B$|$>30$^{\circ}$ and discuss the two groups separately in the following. For brevity, we will call the two groups high- and low-reddening GCs in the following.

\subsection{Cluster-by-cluster search}
\label{sec:ind}

Having concluded that it is unlikely that all our GCs are contaminated by the foreground medium, we are now interested in understanding, as much as possible, which individual GCs are likely to be actually contaminated. We started by examining the catalogs of local molecular clouds within $\simeq$2--3~kpc from the Sun by \citet{zucker20} and \citet{spilker21}, to identify GCs in our sample lying close to well-known cloud complexes. What is especially important is that we do know the distance to the molecular clouds in these two catalogs, so we can be sure that they have nothing to do with the putative ICM of our GCs. We considered the reported cloud sizes and the areas covered by the \citet{stetson19} photometry. In the case of GCs lying just a few degrees from known clouds, we used the SkyView online tool\footnote{\url{https://skyview.gsfc.nasa.gov/}} to better understand the detailed shape of the potentially dangerous molecular clouds. In particular, we focused on the IRAS \citep{neugebauer84}, WISE \citep{wright10}, Akari \citep{murakami07}, and \citet{schlegel98} datasets. In some cases, we noticed large-scale clouds passing over our targets in the 2D maps: we decided to mark these GCs as (potentially) contaminated even if we did not have reliable distance information. 

Additional GCs lie behind high- or intermediate-velocity gas clouds detected by UV spectroscopy \citep{deboer83,welsh12,welsh12b}, namely NGC\,6752, M\,13, and M\,15. In these measurements, the line-of-sight velocity tells that the observed gas is not moving at the same velocity of the GC along the line of sight, and thus it is not within the GCs themselves. However, the column densities of various species in these clouds are so low (on the order of 10$^{12}$--10$^{16}$ atoms per cm$^2$) that they are not expected to contain enough dust to produce appreciable reddening variations, which are detectable with our method. In the case of $\omega$\,Cen, a cloud of 3$\times$10$^{18}$ atoms per cm$^2$ of H\,I was detected with radio observations \citep{smith90} in front of the cluster. Similarly, M\,10, M\,92 and M\,3 lie behind H\,I clouds of 10$^{19}$--10$^{20}$ atoms per cm$^2$ \citep{kerr72,smoker01,howk03,winkel16}, which can potentially impact our DR maps and ICM estimates. This because the typical reddening variations we observe in our maps are on the order of 0.01~mag, which correspond to 10$^{19}$--10$^{20}$ atoms per cm$^2$ \citep{bohlin78,rachford09,liszt14}. 

We also inspected the 3D reddening maps by \citet{green19}\footnote{\url{http://argonaut.rc.fas.harvard.edu/}} and \citet{lallement19,lallement22}\footnote{\url{https://stilism.obspm.fr/}}, which cover distances within $\simeq$3\,kpc from the Sun. We could thus identify GCs lying in regions with several filamentary structures and large cloud complexes, whose distances lie mostly around $\simeq$250\,$\pm$\,75\,pc (range 100--1400\,pc), judging from the cumulative reddening distributions as a function of distance.  These distances are compatible with the expected distribution of Galactic cirri in the Solar vicinity, so whenever we observed a sudden E(B--V) increase, comparable with the typical DR in our reddening maps, we considered the GCs contaminated by foreground cirri. 

Of the 48 GCs in our sample, 33 are contaminated or suspected to be contaminated by foreground medium, four are not obviously contaminated, but belong to the high-reddening sample described in Sect.~\ref{sec:circ}, and finally 11 appear to be clean from any foreground contamination (see also Tab.~\ref{tab:gcs}). It is worth emphasizing at this point that if a GC is contaminated by foreground medium, this does not automatically exclude the presence of an appreciable ICM, it just makes it difficult to separate the ICM (if any) from the foregound contamination. A chief example is M\,3, which has a foreground contamination of about 0.01~mag \citep{howk03,winkel16,lallement19,lallement22}, while our maps show reddening variations of about 0.03~mag. Clearly our DR maps cannot be directly used on M\,3, even if the observed DR cannot be entirely explained by foreground contamination, but we can use this information to attempt a statistical decontamination (see Sect.~\ref{sec:dust}). It is interesting to note that of the 11 clean GCs, 10 are associated to streams of disrupted dwarf galaxies in the Galactic halo \citep{massari19}. On the one hand, this could be due to a selection bias. In fact, high-latitude Galactic regions are less likely to be contaminated by dust, and high-latitude GCs typically move on eccentric orbits that keep memory of their accreted origin. On the other hand, though, accreted GCs are the best candidates to have retained gas and dust within their potential well. In fact, the interactions and shocks they experienced with the Galaxy are less in number (their orbits cross the disk less frequently because of the larger periods) and are less intense due to the dark matter halo of the progenitor galaxy that shields them, at least until its disruption.

   \begin{figure*}[t]
   \centering
   \includegraphics[width=\textwidth]{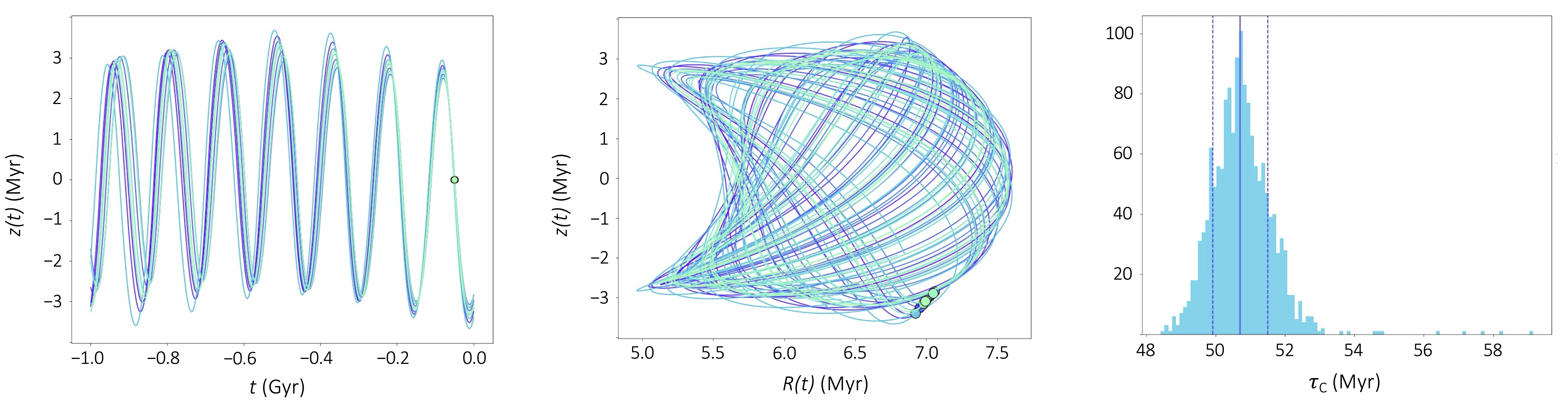}
      \caption{Example of the orbital computation and determination of $\tau_{\rm c}$ for NGC\,104 (47\,Tuc). Variation of the height above the Galactic plane as a function of time (leftmost panel) and galactocentric radius (center panel), for ten randomly selected orbits out of 1500 realizations. The starting point of each orbit integration is plotted as a dot with the same color of the corresponding orbit. The rightmost panel shows the distribution of the 1500 resulting $\tau_{\rm c}$, with its median and MAD as solid and dotted lines, respectively.}
   \label{fig:orbit}
   \end{figure*}


\section{Mass of the intracluster medium}
\label{sec:mass}

In the following, we use our DR maps to estimate the ICM of the sample GCs and we compare them with simple theoretical predictions. 

\subsection{Theoretically expected ICM mass}
\label{sec:theo}

A simple, order-of-magnitude prediction of the expected ICM dust mass was presented by \citet{tayler75} and extensively used in several of the cited studies about the ICM in GCs:

\begin{equation}
    \rm M_{\rm{gas}} = \frac{\tau_{\rm c}}{\tau_{\rm{HB}}}~N_{\rm{HB}}~\mathrm{\delta} \rm M_{\rm{HB}},
\end{equation}

\noindent where $\tau_{\rm c}$ is the time since the last Galactic disk crossing, $\tau_{\rm{HB}}$ the typical lifetime of the horizontal branch (HB) stars, $N_{\rm{HB}}$ the number of HB stars in each GC, and $\delta$M$_{\rm{HB}}$ the typical mass lost by stars before the HB. 

   \begin{figure}[t]
   \centering
   \includegraphics[width=0.99\columnwidth]{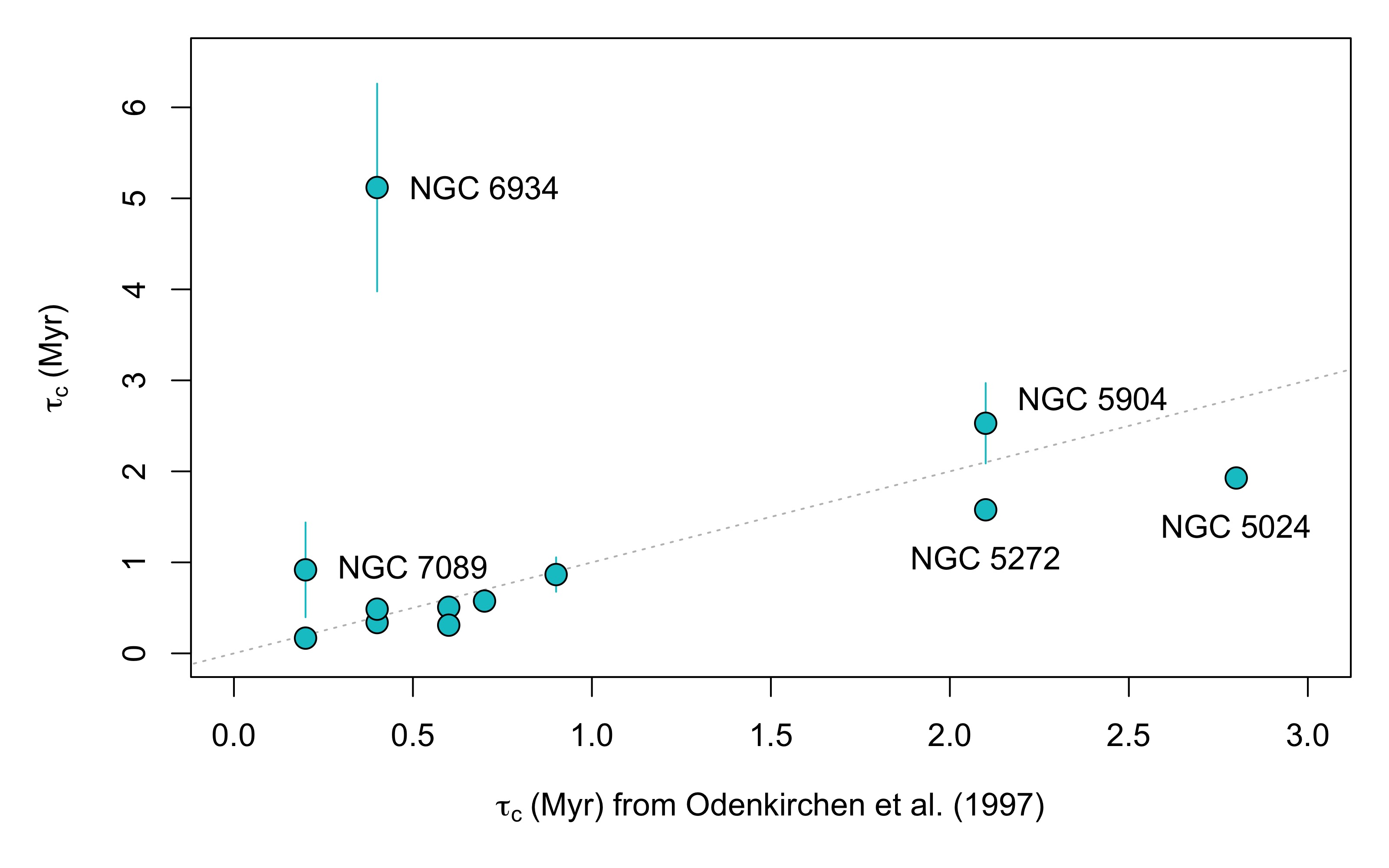}
      \caption{Comparison between our $\tau_{\rm c}$ estimates and those by \citet{odenkirchen97} for 12 GCs in common. Perfect agreement is marked by a dotted line. Errorbars are plotted for each GC, but are mostly smaller than the points. The GCs for which $\tau_{\rm{c}}$ has changed are labeled.}
   \label{fig:tauc}
   \end{figure}

Hydrodynamical studies show that between disk crossings, ram-pressure alone appears unable to remove the entire ICM of a GC \citep{priestley11,chantereau20,naiman20}. Thus, the fundamental assumption is that the entire ICM of the GC is completely cleared during disk crossings. The adopted formulation does not take into account any other dust destruction or removal mechanisms \citep[see, e.g.,][]{umbreit08,chantereau20,pepe16}. We estimated $\tau_{\rm{HB}}$ by interpolating published theoretical HB lifetimes \citep{dorman93} as a function of metallicity, using the [Fe/H] of each GC (Tab.~\ref{tab:gcs}). For N$_{\rm{HB}}$ we used the specific evolutionary flux \citep{renzini83} of the HB phase, which requires the total luminosity of each GC. To compute it, we used integrated colors and magnitudes \citep{harris96,harris10} and bolometric corrections \citep{alonso99}, together with the distance of each GC. We assumed $\delta$M$_{\rm{HB}}$\,=\,0.3\,$\pm$\,0.1\,M$_{\odot}$, which is a commonly accepted value \citep{mcdonald11,buell12,heyl15,tailo21}. 

Estimates of $\tau_{\rm c}$ are only available for 15 GCs \citep{odenkirchen97} and are based on pre-{\em Gaia} orbital computations. We thus used recent proper motions based on Gaia DR3 \citep{gaia_dr3,vasiliev19,vasiliev21} to integrate backward the orbital motion of each GC with a simple but sufficiently accurate model for the Milky Way potential \citep{bovy15}. We performed 1500 orbital realizations, each using a different starting point, extracting the position and velocity of the GC from normal distributions having as mean and variance the current measurements and their uncertainties. Fig.~\ref{fig:orbit} shows an example of $\tau_{\rm c}$ computation. Our final estimates of $\tau_{\rm c}$ for each GC are the medians of the values obtained in the 1500 orbit realizations, with the uncertainties estimated as their MAD (Median Absolute Deviation, Tab.~\ref{tab:gcs}). We compared our $\tau_{\rm c}$ with those by \citet{odenkirchen97}, finding a satisfactory agreement (Fig.~\ref{fig:tauc}).

We also estimated the corresponding dust mass (M$_{\rm{dust}}$) with a simple prescription \citep{draine07,draine11}, which we report in Tab.~\ref{tab:gcs}:

\begin{equation}
\label{eq:dust}
    \frac{\rm M_{\rm{dust}}}{\rm M_{\rm{gas}}} = 0.009 \frac{(\rm O / \rm H)}{(\rm O / \rm H)_{\odot}}.
\end{equation}

The coefficient in Eq.~\ref{eq:dust} changes for different dust models as well as metallicity \citep[e.g.,][]{draine07}. Here we are just interested in an order-of-magnitude estimate and thus we adopt a typical constant dust-to-metal ratio appropriate for the Milky Way \citep[see also Sect.~\ref{sec:dgr} and][]{bianchi19}. The correlation between dust and gas was specifically tested in GCs down to [Fe/H]\,=\,–1.16\, dex \citep{lyons95}, so we extrapolated it to more metal-poor GCs \citep{kahre18}. To account for the metallicity dependence, we used [Fe/H] for each GC from \citet{harris10} and a typical halo [O/Fe]\,=0.3\, dex.

Finally, we estimated the H column density, using two different assumptions on the ICM distribution within each GC, and thus on the area to use for the normalization. In the first hypothesis, we assumed that half of the ICM mass is contained within the half-light radius. This was implicitly or explicitly assumed by the vast majority of the literature studies considered here and it is equivalent to assuming that the ICM follows the stellar density and thus it must be centrally concentrated \citep{angeletti82}. In the second hypothesis, at the opposite extreme, we acknowledge that the pulsar timing data in 47\,Tuc \citep{abbate18} highly disfavor such a concentrated distribution, and thus we considered a homogeneously distributed ICM across the entire GC, within the tidal (truncation) radius. The resulting H column densities are reported in Tab.~\ref{tab:gcs} as th\_N$_{\rm{H}}^{\rm{max}}$ and th\_N$_{\rm{H}}^{\rm{min}}$, respectively. Unsurprisingly, the assumption about the spatial distribution can make the final N$_{\rm{H}}$ vary by one or two orders of magnitude, depending on the GC. If the theoretically expected ICM was centrally concentrated, then the median column density of our GC sample would be N$_{\rm{H}}$\,=\,6.5\,$\times\,$10$^{19}$ cm$^{-2}$ (interquartile range from 1.2\,$\times$\,10$^{19}$ to 2.3$\,\times\,$10$^{20}$), while if the ICM was homogeneously spread over the GC, it would be N$_{\rm{H}}$\,=\,1.1\,$\times\,$10$^{18}$ cm$^{-2}$ (interquartile range from 4.9\,$\times\,$10$^{17}$ to 3.0\,$\times\,$10$^{18}$). 

It should be emphasized that the direct observation of atomic hydrogen through 21 cm emission is not straightforward. \citet{lynch89} tried a deep search ($\simeq$16 hrs) in NGC 6388 that resulted in a non-detection, deriving 3.8 M$_{\odot}$ as an upper limit for the HI mass. Considering a distance of $\simeq$10 kpc, the column density would be $\simeq 1.4 \times 10^{18}$ cm$^{-2}$, below the detection limit of modern HI surveys \citep{peek18}. Furthermore, at 1.42 GHz the beam dilution is not negligible, since the beam size is generally bigger than the GCs angular size, making the detection even more difficult. \citet{faulkner91} searched for HI in NGC 2808, finding atomic hydrogen concentrated near the GC center, with the GC sistemic velocity, but unknown distance. HI is nearly ubiquitous in the Galaxy and it is optically thin, so the full line of sight contributes to the observed profile, and the emission from background and foreground gas may pollute the detection, coupled with stray radiation, complicating the picture. The observed profiles are generally complex, formed by multiple components, hard to associate with environments mapped by the line of sight, despite significant effort in this sense \citep{murray21}.

\subsection{ICM mass estimate from reddening maps}
\label{sec:dust}

For the clean GCs (Tab.~\ref{tab:gcs}) we assume that the amount of foreground dust is negligible, while for the high-reddening or contaminated GCs we assume that the foreground contamination is on average equal to E(B--V) \citep{harris96,harris10}. In both cases, we then use the reddening variations to infer the amount of ICM dust, by assuming that the lowest observed DR corresponds to regions devoid of dust\footnote{This assumption, in the specific case of our contaminated GCs, could lead to an even more overestimated ICM dust mass, on top of the foreground  contamination effect.}, with E(B--V)=0. 

To compute the estimated ICM mass, we used only stars with {\tt flag\,=\,0} in our DR maps (Tab.~\ref{tab:maps}). We further selected stars within r$_{\rm{h}}$, to cover an area as similar as possible to a large fraction of the available upper limits in the literature. We converted the dE(B--V) estimate for each star into a H column density using the relation by \citet{rachford09}

\begin{equation}
    \frac{\rm{N}_{\rm{H}}}{\rm{E(B-V)}}~=~5.94~(\pm 0.34)~\times~10^{21}~\rm{H} ~\rm{cm}^{-2}.
\end{equation}

\begin{table}
\caption{Compilation of the literature ICM estimates in GCs.}
\label{tab:lit}
\centering                         
\begin{tabular}{lcl}        
\hline\hline                
Column              & Units & Description \\
\hline 
Cluster & & Cluster name \\
Mass & (M$_{\odot}$) & Literature mass estimate \\
FoV             & ($^{\prime}$) & Typical field of view size \\
Temperature  & (K) & Assumed dust T (when relevant) \\
Measurement & & Either "detection" or "upper limit" \\
Type & & Whether dust or (ionized) gas \\
Source & & Reference of the source paper \\
\hline
\end{tabular}
\end{table}

Here, we use dE(B--V)--dE(B--V)$_{\rm{min}}$ as our estimate of E(B--V) in each GC, as explained. We further assume that the ICM is dominated by hydrogen. If we were to use any other similar relation in the literature \citep[e.g.,][]{bohlin78} we would obtain results all within $\simeq$5\% from each other. One exception is the \citet{liszt14} relation, which has a coefficient of 8.3 instead of 5.94, implying higher resulting N$_{\rm{H}}$ by up to 30\%. To convert N$_{\rm{H}}$ in H mass, we computed the area patrolled by each star in our DR maps by means of a Voronoi tessellation. This was necessary because the density of stars varies significantly with distance from the GC centers. The masses obtained for the individual areas around each star were then summed to obtain the total H mass. In an attempt to save the high-reddening or foreground contaminated GCs, we corrected the total H mass with the help of the 3D cumulative distributions \citep{lallement19,lallement22}. We first estimated what percentage of the total E(B--V) was reached in the first few kpc from the Sun. We then assumed the same percentage of contamination in the reddening variations, and thus we reduced the total H mass by that percentage. For 13 GCs, the 3D maps provided a higher E(B--V) than the \citet{harris96,harris10} catalog, and thus we did not attempt to estimate the ICM because it was not possible to compute a meaningful correction. 

For convenience, we also computed the dust mass and the median N$_{\rm{H}}$, which are reported in Tab.~\ref{tab:gcs} along with the gas mass estimates. For the dust mass, we used Eq.~\ref{eq:dust} and the same assumptions employed in Sect.~\ref{sec:theo}. We finally computed the N$_{\rm{H}}$ for each GC, as the weighted median (and weighted MAD) of the N$_{\rm{H}}$ estimates obtained for each star in the map (within r$_{\rm{h}}$), and we used the Voronoi area around each star as the weights (see Tab.~\ref{tab:gcs}).

\subsection{Results}
\label{sec:res}

\begin{figure}[t]
   \centering
   \includegraphics[width=0.99\columnwidth]{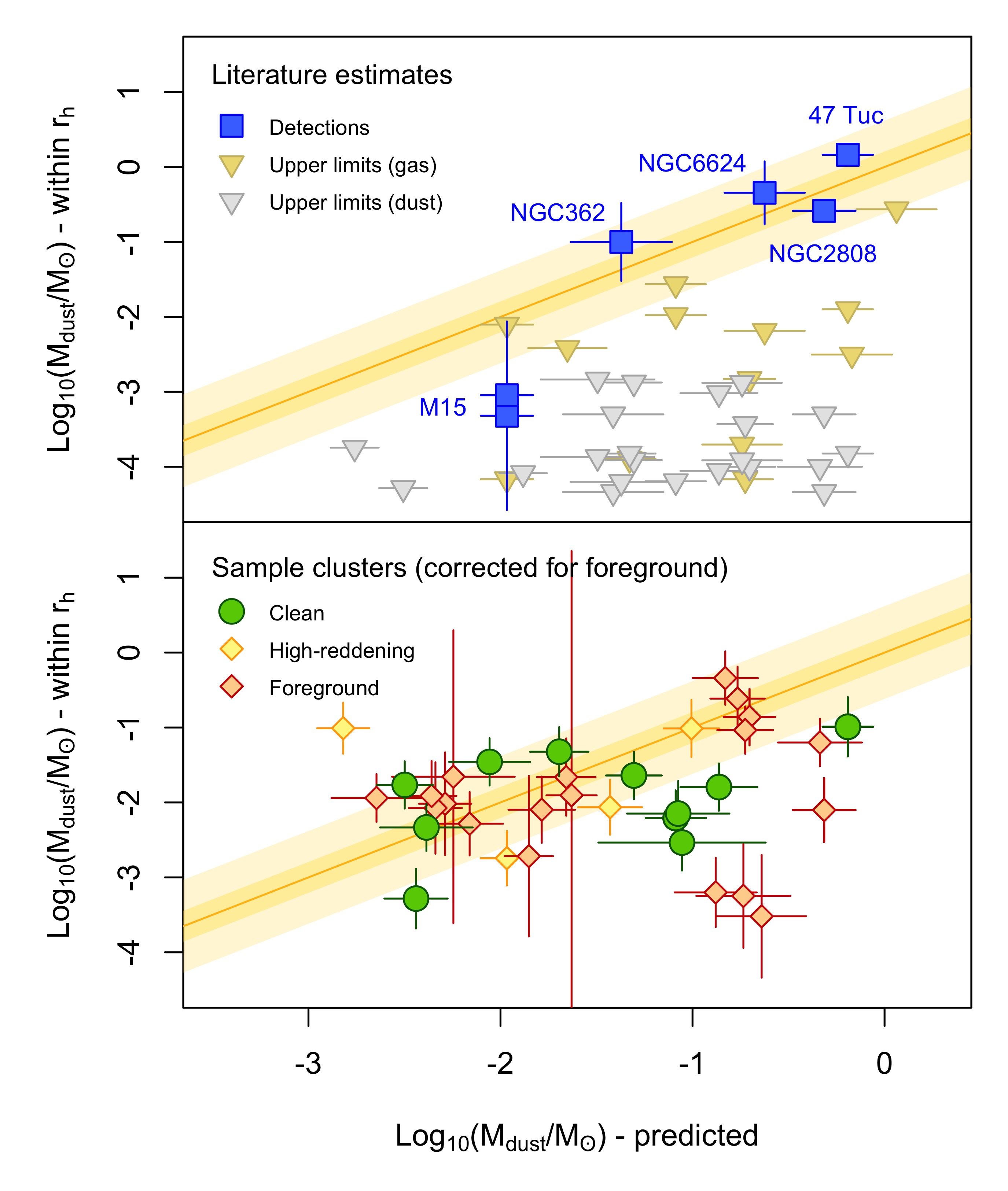}
      \caption{{\em Top panel}. Comparison between the literature upper limits (triangles) and detections of dust in GCs (squares, see Tab.~\ref{tab:lit}) with the theoretical expectations from Tab.~\ref{tab:gcs}. Perfect agreement is marked by the orange line, where 1 and 3\,$\sigma$ uncertainties are shaded in yellow and dark yellow, respectively. {\em Bottom panel}. Our dust estimates for the \citet{stetson19} GCs are compared with the same theoretical predictions. The clean GCs are plotted in green, the high-reddening ones in yellow, the contaminated in red (see Sect.~\ref{sec:maps}). All estimates were corrected for foreground contamination using 3D cumulative reddening maps \citep{lallement22}. 1\,$\sigma$ uncertainties are plotted in all panels.}
   \label{fig:res}
   \end{figure}

Our ICM mass estimates, once corrected for the foreground contribution, agree with theoretical expectations, within a 3\,$\sigma$ spread of about one order of magnitude, as illustrated in Fig.~\ref{fig:res}. There is a tail of low dust mass estimates, especially for the high-reddening GCs (red symbols in Fig.~\ref{fig:res}, bottom panel), which could be a real feature, for example because most of these GCs are in regions with a higher density Galactic medium, so they could suffer from higher degrees of ram-pressure stripping. However, it could also be the result of our foreground correction procedure. The median difference and MAD, in the sense of our dust mass observations minus predictions, is $-$0.02\,$\pm$\,0.07\,M$_{\rm{\odot}}$ (Fig.~\ref{fig:res}, bottom panel). 

Our estimates agree also with previous literature detections, which in turn agree with theoretical predictions, as already mentioned in Sect.~\ref{sec:intro} (see Fig.~\ref{fig:res}, top panel). The median difference and MAD between literature detections and theoretical expectations is 0.02\,$\pm$\,0.17\,M$_{\rm{\odot}}$. The median difference and MAD between our dust mass estimates and the literature detections is $-$0.13\,$\pm$\,0.19\,M$_{\rm{\odot}}$, based on 47\,Tuc, NGC\,2808, and M\,15. It is worth noting that the two detections of the ICM in M\,15 \citep{evans03,boyer06}, shown as the blue squares on the lower-left side of Fig.~\ref{fig:res} (top panel), are lower than theoretical predictions, but if the assumptions on dust type and temperature change, the ICM mass would change by about one order of magnitude, according to \citet[][see also next section]{evans03}.

Upper limits from the literature, on the other hand (downward triangles in Fig.~\ref{fig:res}, top panel), are several orders of magnitude lower than literature detections, theoretical expectations, and our estimates. We note that the upper limits obtained by studying the gas emission (HI or molecules) are often one or two orders of magnitude higher than the ones obtained by studying the IR dust emission. It appears thus clear that the long-standing problem of the missing ICM in GCs is confined to upper limits derivations, and thus it may be a false problem.


\section{Discussion}
\label{sec:disc}

In the following, we discuss possible reasons why the upper limits in the literature may have been underestimated.

\subsection{ICM spatial distribution}
\label{sec:spat}

All the literature studies mentioned above assume that the ICM distribution follows the stellar distribution \citep[see, e.g.,][]{angeletti82}, and therefore one expects roughly that half of the ICM mass is contained within r$_{\rm{h}}$. However, as already mentioned (Sect.~\ref{sec:intro} and \ref{sec:theo}), the pulsar timing data by \citet{abbate18} highly disfavor such a concentrated distribution and suggest that a better fit would be a flatter or uniform distribution. Indeed, when computing the expected N$_{\rm{H}}$ in Sect.~\ref{sec:theo}, we noted that if the gas was concentrated in the GC center, then we would expect to have a detectable amount of neutral hydrogen ($\sim$10$^{20}$~cm$^2$), while if the gas is spread over the entire extent of the GC, the column density drops by about one to two orders of magnitude, depending on the GC. Additionally, in the DR maps of our clean GCs we only rarely observe central concentrations of reddening: structures such as bubbles, stripes, or rings are more common (see Figs.~\ref{fig:allmaps1}--\ref{fig:allmaps4}). 

If the ICM distribution is indeed flatter than assumed, this would have an important impact on the expected column densities, making the molecular and atomic gas up to 100 times more difficult to detect and the upper limits on the ICM mass underestimated. In fact, the assumption that the beam, by covering the GC center, samples the entirety of the ICM leads to an underestimate of the upper limit, compared to the case in which one assumes to sample only a (small) part of the ICM. The beam size, combined with the assumption that the ICM is concentrated, would have a similar impact on IR observations to detect the dust emission (see Tab.~\ref{tab:lit}). Additionally, if a beam away from the GC center is used to estimate the foreground dust emission, and if this beam is still within the area covered by the ICM, then the final upper limit would be even more underestimated \citep[e.g., see][]{faulkner91}. Of course, there could also be intermediate situations in which the ICM is still somewhat concentrated to the center, but with a flatter distribution than that of the stars, so the case of a completely uniform ICM distribution should be seen as a limiting case, at least until more data are available.

In conclusion, the assumption of a concentrated ICM is one of the main causes for the underestimate of the literature ICM mass upper limits, accounting for one or even two orders of magnitude, depending on the actual beam sizes of each study and on the true ICM distribution. Removing this assumption would be enough to reconcile most of the gas upper limits with theoretical expectations (gold triangles in Fig.~\ref{fig:res}) and would significantly alleviate the discrepancy for the dust upper limits (silver triangles in Fig.~\ref{fig:res}).

\subsection{Differences in dust-to-gas ratio or ionization}
\label{sec:dgr}

Another reason why the upper limits could be difficult to estimate correctly, is that the ICM could be different from the Galactic medium as far as the dust-to-gas ratio (DGR), the ionization fraction, or the relative amount of molecular gas are concerned. 

The DGR is an essential property of the gaseous medium, both in the Galactic disk and in GCs and its value heavily alters the thermal, chemical and dynamical state of the medium. There are several studies of the DGR in molecular clouds in our Galaxy \citep[see][]{reach17,monaci22} and in other galaxies \citep[see][]{lisenfeld98,li19,devis19,remy14}. Even if there is a widely adopted value for the DGR \citep[100 -- 150, see][]{hildebrand83}, it is well known that the ratio changes between galaxies \citep{young86} and even within our Galaxy \citep{reach15}. 

It should be pointed out that the DGR and the dust-to-metals ration (D/Z) can, in principle, be different in GCs: in that environment, dust is almost entirely created in stellar winds and is exposed to a different radiation field with respect to the diffuse interstellar medium. Considering a stronger radiation field in GCs, the dust grains may be destroyed more efficiently, changing the DGR. Since the molecular gas is formed onto dust grains surface, the radiation field in GCs can dismantle the molecules, increasing the relative gas abundance. Despite the several studies about the DGR and D/Z in galaxies at different $z$ and their evolution \citep[see][]{li19, choban22, parente22}, we did not find any study dedicated specifically to the DGR in GCs, except for surveys in the 70's to 90's \citep{knapp74,mirabel79,knapp96}, which generally found a DGR in GCs comparable with those determined along other lines of sight. Among the cited literature sources, we indeed found no evidence that GCs should possess a different ratio, while we found confirmations that it should be similar to what expected \citep[see, e.g.,][]{mcdonald09}. Thus, we use the customary DGR and we do not think that this specific quantity mislead previous estimates of the ICM upper limits.

Dust plays a crucial role also in the formation of molecular gas. The formation of molecular hydrogen in the gas-phase is extremely inefficient \citep{leteuff00}, whilst the formation on dust surfaces was first discussed by \citet{gould63} and it is much more efficient. However, the formation rate of H$_2$ depends on the dust and gas temperature and density: in the past years several studies were performed, both theoretically \citep[see][] {burke83,buch91} and experimentally \citep[see][]{pirronello97b,pirronello97a} and they show that the H$_2$ formation is more efficient at low temperatures (i.e., in the Cold Neutral Medium is enhanced), therefore the formation of molecular hydrogen in GCs might be quenched if the ICM is really warmer than in the Galactic medium, as assumed in many ICM searches. This would make the ICM detection through molecular tracers such as CO, CH and OH difficult. Additionally, in GCs the molecules are more exposed to the stellar radiation which can break them up. It is then possible that the lack of molecular emission in GCs is partially due to the chemical and thermal state of the ICM. For example, \citet{leon96} searched for $^{12}$CO emission in six different GCs, using the IRAM 30-m radio telescope, down to 0.005 K but they did not detect any molecular gas even if it should have been there according to theoretical expectations. Even the formation of CO in the gas-phase is possible, the predominant production mechanism is on the dust grains surface, therefore the CO relative abundance is strongly dependent on the adopted dust model \citep[see][]{draine11}. Furthermore, dust segregation (by grain size) can also be possible in GCs, since the stars may change the spatial distribution of the dust through stellar winds and radiation. According to the simulation by \citet{umbreit08}, grains with size $\simeq\,0.01\,\mu m$ are destroyed, those of $\simeq\,0.1\,\mu m$ are dragged by gas dynamics, and those of $>1\,\mu m$ tend to be segregated in the GC center. It is then possible that the CO production is different in GCs, with regions where the molecular gas is enhanced and other regions were it is depleted, making the detection even more difficult.

Finally, concerning the fraction of ionized to neutral gas, we bring as an example the study by \citet{mcdonald15}, who estimated the ionizing effect of all the blue stars in GCs, especially white dwarfs, and concluded that most if not all the ICM should be constantly and fully ionized. If correct, this would not only offer an explanation on why there are so few detections of neutral hydrogen in GCs, but also to the problem of the spatial distribution of the ICM in GCs. In fact, a pressure supported gas would expand within the GC and its distribution would not be centrally concentrated. Additionally, a significant fraction of the ICM would reach the GC limit and escape.

\subsection{Dust composition}
\label{sec:comp}

It appears now ascertained that the dust produced around asymptotic and red giant branch stars in GCs contains a sizeable fraction of metallic iron grains \citep{mcdonald10}. This is confirmed by \citet{marini19}, who studied the metal-poor environment around asymptotic red giants in the Large Magellanic Cloud, and found solid iron grains. Metallic iron grains are indeed included in recent modeling of dust evolution \citep{choban22,choban24}. The production of iron dust is thought to become important at low metallicity, below [Fe/H]\,$\simeq$\,$-$1\,dex. Below the limit of GC metallicities, at [Fe/H]\,$\lesssim$\,$-$2.5\,dex, theoretical studies predict a predominance of carbonaceous dust \citep{ventura21}. Additionally, the formation of iron dust in the GC regime would explain several other observational facts, including the depletion of Fe in the interstellar medium \citep{mcdonald10,zhukovska18}. Because dust in GCs is continuously replenished by asymptotic red giants, this contradicts the assumption of pure silicate dust grains, used in many of the literature studies cited throughout the text and in Tab.~\ref{tab:lit}. For example, \citet{evans03} and \citet{boyer06} assume fayalite grains and a typical Milky Way two-component model, respectively. They stated that uncertainties in the dust temperature and composition would imply an order-of-magnitude extra uncertainty in their ICM estimate for M\,15 (see also Fig.~\ref{fig:res}). 

Changing the dust model changes several parameters, including the grain size, the equilibrium temperature, the spectral energy distribution of the dust reemission, and renders all the upper limits (and the M\,15 detections) more uncertain. In particular, iron dust grains should be smaller than silicate grains, and they should heat and cool much faster. Thus, their actual temperatures could vary a lot around a typical temperature, which should be lower. Certainly, all the previous (before 2010) upper limits on the ICM dust mass based on the assumption of silicate grains should be revised.  While a detailed discussion of the ICM estimate uncertainties associated to the dust composition is out of the scope of the present work, we can follow \citet{evans03} and assume that it should be of at least one order of magnitude. If that is correct, then the discrepancy with theoretical expectations could be further reduced by another order of magnitude, on top of the considerations on the ICM spatial distribution (Sect.~\ref{sec:spat}). 

\subsection{Dust temperature}
\label{sec:temp}

Most of the upper limits derived in the IR, looking for dust reemission, assumed a dust temperature higher than the typical 20\,K average temperature of Galactic dust, ranging from 35 to 70\,K (see Tab.~\ref{tab:lit}). This has a direct implication on the ICM or dust mass derived from the IR observational upper limits. For example, according to the computations by \citet[][Tab.~5]{hopwood99}, the assumption of a temperature lower by about 30\,K would imply an increase of the estimated dust mass by about one order of magnitude, significantly decreasing the tension between upper limits and theoretical expectations, especially if combined with the assumpions on the ICM distribution within GCs (Sect.~\ref{sec:spat}). Such choices of the equilibrium temperatures were only partially motivated by computations such as those by \citet{angeletti82}, assuming pure graphite ($\simeq$70--80\,K) or silicate grains ($\simeq$40\,K). Even setting aside considerations related to dust composition (Sect.~\ref{sec:comp}), some of the temperatures assumed in the literature appear to be high compared to those assumed by other studies in the case of silicate dust models. For example, the computation by \citet{evans03} assumed 70\,K for fayalite in M\,15, while perhaps values around 40--50\,K would have been more appropriate \citep{angeletti82}, increasing their mass estimate by one order of magnitude. 



\section{Conclusions}
\label{sec:concl}

We have used the wide-field, ground-based photometry of 48 GCs \citep{stetson19} to derive extremely detailed differential reddening maps, using classical techniques. The maps can be used to remove the DR effects from the photometry and in particular from CMDs. Thanks to a detailed quality assessment, the maps, or their cleanest portions, can also be used to correct other photometries, not just the \citet{stetson19} ones from which they were obtained. 

We used the maps to estimate the ICM mass of the sample clusters. We found 11 of our GCs to be essentially free from foregound contamination, 33 to be contaminated or potentially contaminated to various degrees, and 4 to be still free from contamination, but located too close to the Galactic plane. We applied a correction for the foreground contamination with the help of 3D reddening maps \citep{lallement19,lallement22}. Our final ICM mass estimates are compatible with the theoretical expectations outlined by \citet{tayler75}, within approximately one order of magnitude, and they agree well with the few ICM detections (Sect.~\ref{sec:res}).

It appears thus likely that the long-standing discrepancy between observations and expectations is driven entirely by an underestimation of the upper limits, and we discuss some of the reasons why they could be underestimated (Sect.~\ref{sec:disc}).

After carefully considering recent evidence as well as our own results, we conclude that: 
\begin{enumerate}
    \item{the ICM distribution is unlikely to follow that of the stars, but should be flatter or even uniform \citep[Sect.~\ref{sec:spat}, see also][]{abbate18}; our DR maps rarely show central concentrations of reddening, supporting the findings by \citet{abbate18} for 47\,Tuc. This is sufficient to reconcile most gas upper limits with expectations;} 
    \item{the dust composition in GCs may be dominated by iron grains and not by silicates, compromising the literature derivation of ICM mass upper limits from non-detections \citep[][]{mcdonald10,marini19}. The derivation of upper limits from IR non-detections, obtained before 2010, should thus be revised (see Sect.~\ref{sec:comp} and \ref{sec:temp}).}
\end{enumerate}

The combination of these two new facts is sufficient to increase the ICM mass estimate associated to IR dust emission upper limits and of radio searches for neutral gas, as well as the M\,15 detections \citep{evans03,boyer06}, by several orders of magnitude, thus potentially reconciling the upper limits with theoretical expectations for dust searches as well. 

We conclude that, based on current evidence, including our own results, there is no discrepancy between the observed and expected ICM masses in GCs. The puzzling and long standing lack of ICM in GCs is thus a false problem. The misunderstanding was induced by our past (before 2010) insufficient knowledge of the characteristics and spatial distribution of dust in GCs, which caused an underestimate of the ICM mass upper limits.


\begin{acknowledgements}

{\bf People.} We would like to thank the following colleagues for discussions, suggestions, and ideas: Giuseppe Altavilla, Angelo L. Antonelli, Ivan Cabrera-Ziri, Claudio Codella, Rodrigo Contreras, Bruce Draine, Michele Fabrizio, Davide Fedele, Antonio Garufi, Silvia Marinoni, Paola Marrese, Raul M. Murillo, Linda Podio, and Manuela Zoccali.\\

{\bf Funding.} This project has received funding from the European Research Council (ERC) under the
European Union’s Horizon Europe programme {\em "StarDance: the non-canonical evolution of stars in clusters"} (ERC-2022-AdG, Grant Agreement 101093572, PI: E. Pancino). We also acknowledge the financial support to this research by INAF, through the Mainstream Grant {\em “Chemo-dynamics of globular clusters: the Gaia revolution”} (n. 1.05.01.86.22, P.I. E. Pancino). EP and DM  acknowledge funding by the European Union – NextGenerationEU" RRF M4C2 1.1  n: 2022HY2NSX. "CHRONOS: adjusting the clock(s) to unveil the CHRONO-chemo-dynamical Structure of the Galaxy” (PI: S. Cassisi). C.E.M.-V. is supported by the international Gemini Observatory, a program of NSF's NOIRLab, which is managed by the Association of Universities for Research in Astronomy (AURA) under a cooperative agreement with the National Science Foundation, on behalf of the Gemini partnership of Argentina, Brazil, Canada, Chile, the Republic of Korea, and the United States of America. \\

{\bf Data.} This work has made use of data from the European Space Agency (ESA) mission Gaia (\url{https://www.cosmos.esa.int/gaia}), processed by the Gaia Data Processing and Analysis Consortium (DPAC, \url{https://www.cosmos.esa.int/web/gaia/dpac/consortium}). Funding for the DPAC has been provided by national institutions, in particular the institutions participating in the Gaia Multilateral Agreement.\\

{\bf Tools.} We extensively used TopCat for data visualization and exploration \citep{taylor05}. All figures and most of the data analysis were made with the R programming language \citep{R} and Rstudio (\url{https://www.rstudio.com/}) or with Python (\url{https://www.python.org/}) and
specifically with the packages numpy, scipy, and matplotlib. 

\end{acknowledgements}


\bibliographystyle{aa} 
\bibliography{reddening} 


\appendix

\section{Reddening maps}
\label{sec:allmaps}

\begin{figure*}[t]
   \centering
   \includegraphics[width=\textwidth]{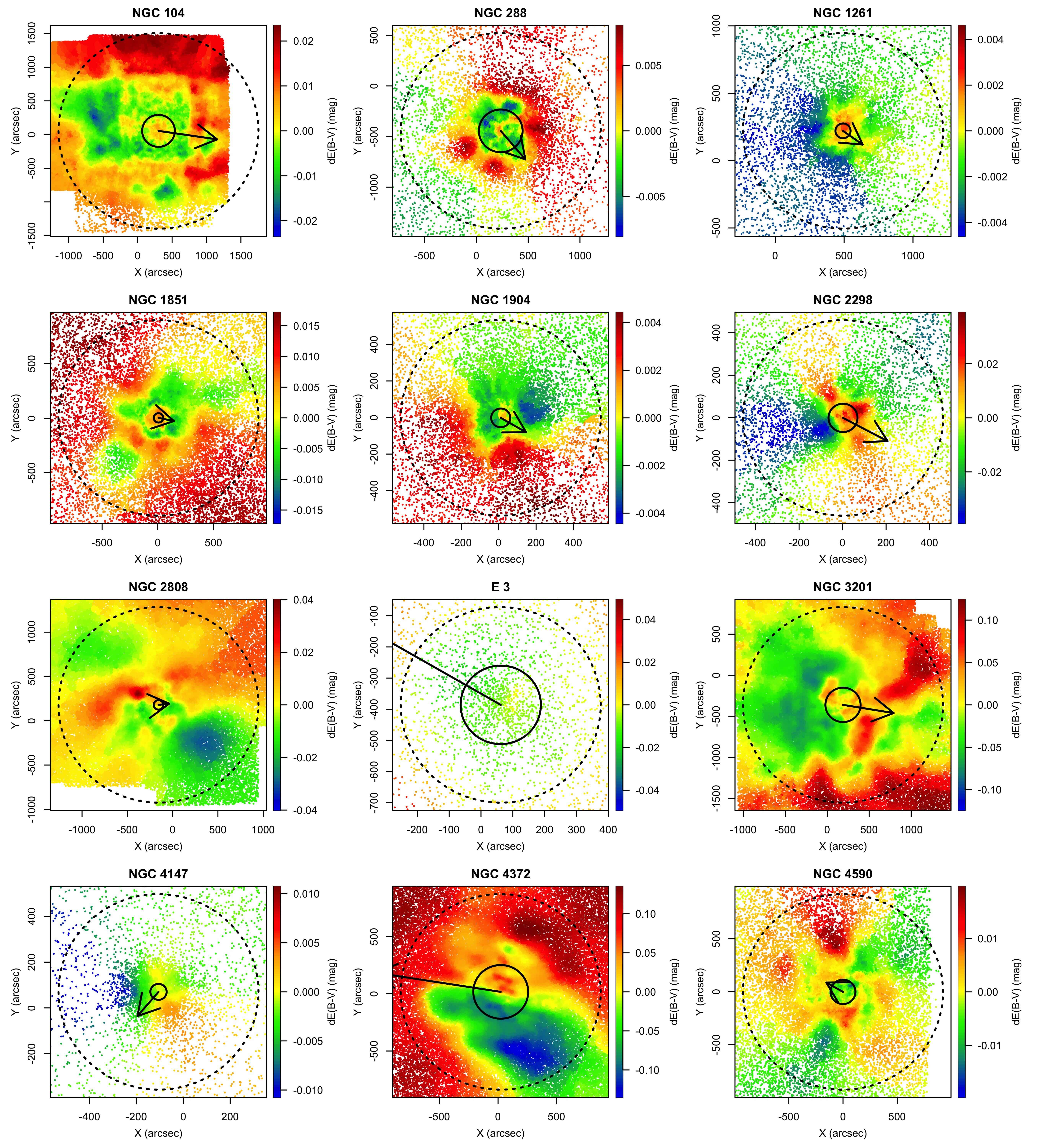}
      \caption{Differential reddening maps for the first twelve GCs, alphabetically sorted. See Appendix~\ref{sec:allmaps} for details.}
   \label{fig:allmaps1}
   \end{figure*}

\begin{figure*}[t]
   \centering
   \includegraphics[width=\textwidth]{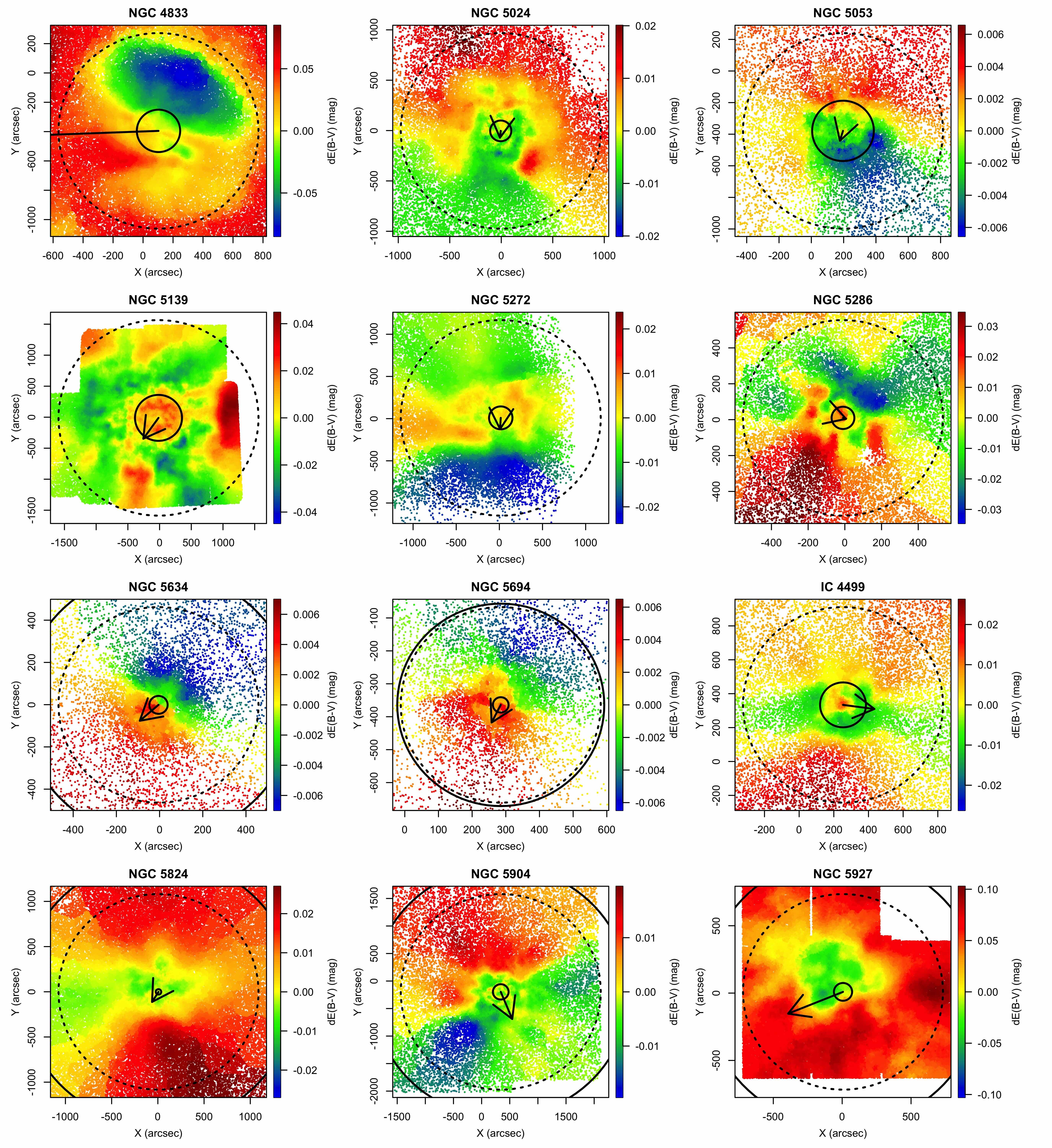}
      \caption{Same as Fig.~\ref{fig:allmaps1}, but for the second set of twelve GCs.}
   \label{fig:allmaps2}
   \end{figure*}

\begin{figure*}[t]
   \centering
   \includegraphics[width=\textwidth]{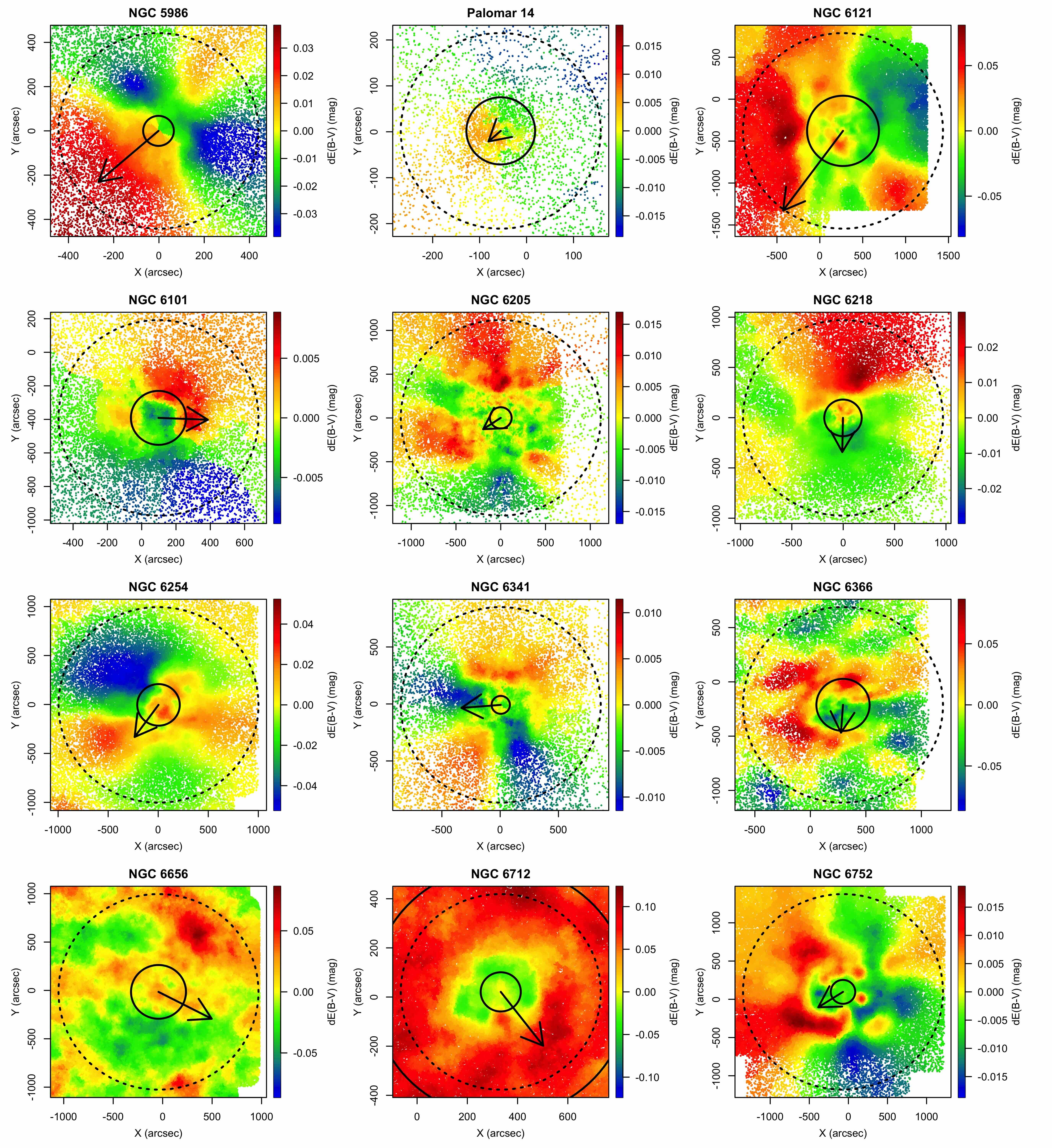}
      \caption{Same as Fig.~\ref{fig:allmaps2}, but for the thirds set of twelve GCs.}
   \label{fig:allmaps3}
   \end{figure*}

\begin{figure*}[t]
   \centering
   \includegraphics[width=\textwidth]{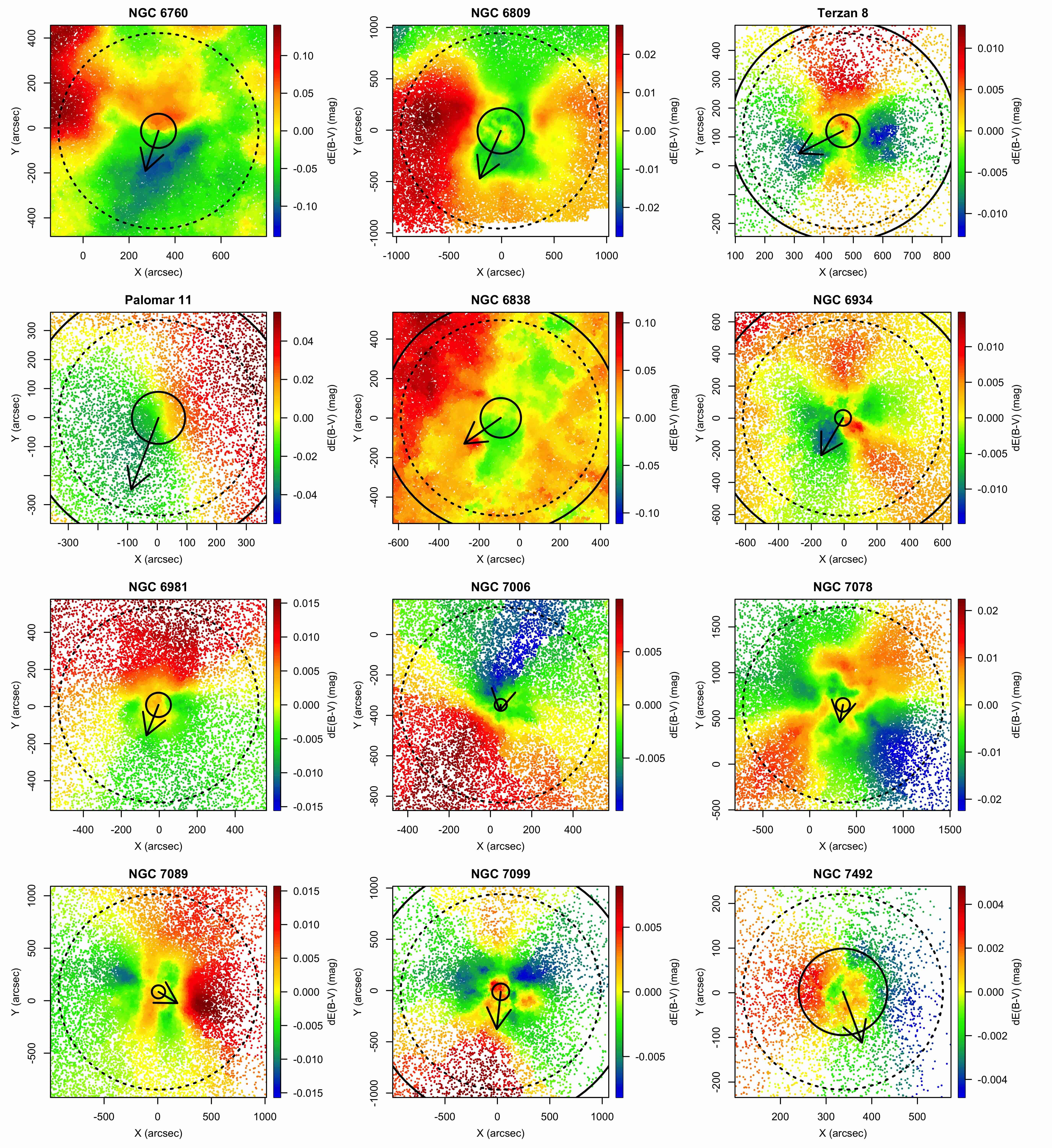}
      \caption{Same as Fig.~\ref{fig:allmaps3}, but for the last set of twelve GCs.}
   \label{fig:allmaps4}
   \end{figure*}

Figs.~\ref{fig:allmaps1}, \ref{fig:allmaps2}, \ref{fig:allmaps3}, and \ref{fig:allmaps4} show the differential reddening maps for the 48 GC photometric catalogs by \citet{stetson19}. Each map is color coded with the exact DR value in Tab.~\ref{tab:maps} and is centered on zero, which is the map median reddening value, which is color coded in yellow. The half light radius r$_{\rm{h}}$ and tidal or truncation radius r$_{\rm{t}}$, when visible, are marked with a small and large solid circle, respectively. The limiting radius r$_{\rm{\ell}}$, beyond which the maps are not significant anymore, is marked with a dotted circle. Each panel reports the GC proper motion as an arrow \citep{vasiliev19}.


\end{document}